\documentclass[11pt,a4paper]{article} \usepackage{jheppub}
\usepackage{graphicx}

\usepackage{epsfig}
\usepackage{color}
\usepackage{amssymb}

\begin{document}
\title{On the ratio of  $\boldsymbol{t\bar{t}b\bar{b}}$
and $\boldsymbol{t\bar{t}jj}$  cross sections at the CERN Large Hadron 
Collider}

\author{G. Bevilacqua$^{a}$ and  M. Worek$^{b}$}
\affiliation{
$^{a}$ INFN, Laboratori Nazionali di Frascati,
Via E. Fermi 40, I-00044 Frascati, Italy}
\affiliation{$^{b}$ Institute for Theoretical Particle Physics and Cosmology, 
RWTH Aachen University, \\Otto-Blumenthal Str., D-52056 Aachen, Germany} 

\emailAdd{Giuseppe.Bevilacqua@lnf.infn.it}
\emailAdd{worek@physik.rwth-aachen.de} 

\abstract{Triggered by ongoing experimental analyses, we report on a
  study of the cross section ratio $\sigma(pp \to
  t\bar{t}b\bar{b})/\sigma(pp \to t\bar{t}jj)$ at the next-to-leading
  order in QCD, focusing on both present and future collider energies:
  $\sqrt{s}=$ 7, 8, 13 TeV. In particular, we provide a comparison
  between our predictions and the currently available CMS data for the
  8 TeV run. We further analyse the kinematics and scale uncertainties
  of the two processes for a single set of parton distribution
  functions, with the goal of assessing possible correlations that 
  might help to reduce the theoretical error of the ratio and thus
  enhance the predictive power of this observable. We argue that the
  different jet kinematics makes the $t\bar{t}b\bar{b}$ and
  $t\bar{t}jj$ processes uncorrelated in several observables, and show
  that the scale uncertainty is not significantly reduced when taking
  the ratio of the cross sections.}

\dedicated{\rm TTK-14-03}

\maketitle

%
\section{Introduction}
In order to establish whether the scalar resonance observed at the
Large Hadron Collider (LHC) around 125 GeV
\cite{Aad:2012tfa,Chatrchyan:2012ufa} matches the properties of the
Standard Model (SM) Higgs boson, quantities such as the couplings to
fermions have to be measured with high precision. A special interest
is due to the Yukawa couplings to top ($Y_{t}$) and bottom
($Y_{b}$) quarks. Massive as they are, these quarks are ideal
candidates for probing the nature of the new particle and more
generally of the Electroweak Symmetry Breaking mechanism.

For a SM Higgs boson with the observed mass value, the dominant decay
mode  is $H\to b\bar{b}$ \cite{Djouadi:2005gi}. The presence of an
overwhelming QCD background discourages Higgs searches in the direct
production channel $pp \to H \to b\bar{b}$. Attention is rather put on
Higgs production in association with one or more additional objects
\cite{Aad:2012gxa,Chatrchyan:2012ww,Chatrchyan:2013yea,Chatrchyan:2013zna}
due to the fact that backgrounds are easier to control in such an
environment. 

Among all the associated production mechanisms that have
been explored by the ATLAS and CMS Collaborations,  the $pp\to
t\bar{t} H \to t\bar{t}b\bar{b}$  channel plays an important role 
\cite{Desch:2004kf,Lafaye:2009vr,Klute:2012pu}. The
production rate for  this process is directly sensitive to $\sim
(Y^2_{t} \, Y^2_{b})/\Gamma_{H}$, where $Y_{t}$  and  $Y_{b}$ are the
top- and the bottom-Yukawa coupling respectively and  $\Gamma_{H}$ is
the Higgs boson width (see Figure 1, diagram A). 
Since the total Higgs boson width can be constrained via independent
measurements, {\it e.g.} by the ratio of off-shell and on-shell
production  and decay rates in the $H\to ZZ \to 4\ell$ and/or the   
$H\to W^{+}W^{-} \to 2\ell \,2\nu $ 
channel \cite{Caola:2013yja,Campbell:2013una,Campbell:2013wga,CMS5},  
the $t\bar{t}H\to t\bar{t}b\bar{b}$ process adds to the information on
$Y_{b}$ provided  by  $pp\to VH (H \to b\bar{b})$, where
$V=Z/W^{\pm}$.

\begin{figure}
\begin{center}
\includegraphics[width=0.95\textwidth]{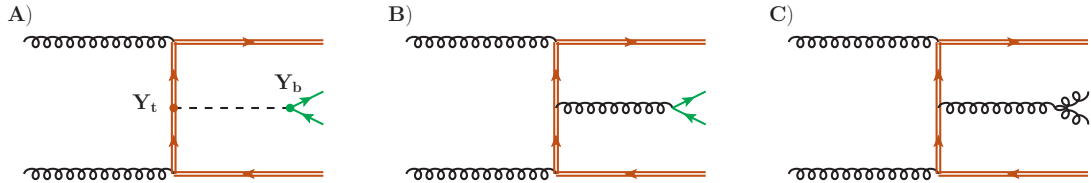}
\end{center}
\caption{\it \label{signal} Representative Feynman diagram for the
associated production of the Higgs boson and a $t\bar{t}$ pair
followed by the decay of the Higgs boson into a $b\bar{b}$ pair
(diagram A). Also shown are representative Feynman diagrams for  the
irreducible background with the same final state (diagram B), as well
as the reducible  background with two jets (diagram C). A single
dashed line corresponds to the Higgs boson, double lines correspond to
top quarks, single lines to bottom quarks and wiggly ones to gluons.}
\end{figure}

However, the $t\bar{t}H(H\to b\bar{b})$ final state is very
challenging to measure. Search strategies employed by both experiments
are based on the full reconstruction of the $t\bar{t}b\bar{b}$ final
state from charged leptons, missing energy and jets
\cite{Drollinger:2001ym,ATLAS,CMS1,CMS2}.  Using $b$-jet tagging,
events with four $b$-jets are isolated, and the decays of the two
candidate top quarks are reconstructed. Afterwards, the two $b$-jets
which have not been associated to top decays are assigned to the
candidate Higgs boson's decay. It should be clear that the
identification of such decay products is not free of ambiguities. The
so-called combinatorial background is responsible for a substantial
smearing of the Higgs boson peak in the $b\bar{b}$ invariant
mass. Together with the possibility of  misidentifying light jets with
$b$-jets, this represents a serious obstacle to the observation of the
Higgs signal and demands a good control of dominant backgrounds as a
prerequisite for a successful analysis.  Several
strategies have been presented in the literature  to increase a
sensitivity for this challenging channel. The most promising being the
jet substructure  techniques  for  boosted heavy states and the matrix
element  methods \cite{Plehn:2009rk,Artoisenet:2013vfa,Bramante:2014gda} 
to name just a few examples. 

The process of $t\bar{t}b\bar{b}$ production in QCD is the most
important irreducible background for the signal under
consideration (see Figure \ref{signal}, diagram B). 
With the help of $b$-jet tagging algorithms, it is
possible to isolate the contribution of this process from the most
general reducible background represented by $t\bar{t}jj$
production (see Figure \ref{signal}, diagram C).
 Instead of extracting absolute cross sections, one can
measure the production rate of $t\bar{t}b\bar{b}$ normalized to the
inclusive $t\bar{t}jj$ sample. This procedure has been explored by
both CMS and ATLAS Collaborations \cite{CMS3,CMS4,Aad:2013tua} and has
the advantage that many experimental systematics, including luminosity
uncertainty, lepton identification and jet reconstruction efficiency,
are expected to cancel in the ratio. The overall  systematic error
should thus be dominated by  the efficient and clean identification of
bottom jets, referred to as the b-jet tagging efficiency, as well as the
tagging efficiency for the light flavor jets, referred to as the
mistag rate.

On the theory side, the QCD backgrounds $pp(p\bar{p})\rightarrow
t\bar{t}b\bar{b}$ and $pp(p\bar{p})\rightarrow t\bar{t}jj$ have been
calculated at the next-to-leading order (NLO) in QCD
\cite{Bredenstein:2009aj,Bevilacqua:2009zn,
  Bredenstein:2010rs,Worek:2011rd,
  Bevilacqua:2010ve,Bevilacqua:2011aa}.    Fairly moderate,
$\mathcal{O}(15\%-30\%)$ corrections have been found for both
processes. The estimated theoretical uncertainties due to truncation
of higher-order terms in the perturbative expansion are of the same
size.  In addition, first results for $t\bar{t}$ production in
association with either two light or two bottom jets, and enhanced
by a parton shower have recently appeared
\cite{Hoeche:2014qda,Cascioli:2013era,Kardos:2013vxa}. Scale
variations before and after matching have been assessed to be rather
similar.  Each of these calculations, however, has been carried out
with different sets of cuts, jet algorithms, values of top quark mass
and parton distribution functions (PDFs). This makes a determination
of the cross section ratio possible only at the price of introducing
undesired additional theoretical uncertainties.

The purpose of this paper is twofold. First, we would like to provide
a systematic analysis of $t\bar{t}b\bar{b}$ and $t\bar{t}jj$
backgrounds and extract the most accurate NLO predictions for the
cross section ratio, to be used in comparisons with the available LHC
data. The second goal is to examine whether the ratio has enhanced
predictive power for Higgs searches, by investigating possible
correlations between the two processes in the quest of reducing
theoretical errors.

The paper is structured as follows. In Section \ref{framework} we
assess the kinematical range of our predictions, {\it i.e.}  we
motivate which phase space restrictions, particularly in the
transverse momentum of jets, shall be applied for our fixed-order
results to be reliable. Beyond these limits, the stability of the
perturbative expansion is likely to be endangered, and resummation of
higher order effects is required. We estimate these limits by studying
leading-order $t\bar{t}jj$ production matched with \textsc{Pythia}
parton shower, and use the obtained results to determine the kinematical
setup for our predictions. In Section \ref{nlo-distributions} we
examine next-to-leading order differential cross sections for both
$t\bar{t}b\bar{b}$ and $t\bar{t}jj$ processes, analysing  similarities
and possible correlations between the two backgrounds. Subsequently,
we provide in Section \ref{nlo-ratio} the results for the ratio and
absolute cross sections for three different collider energies:
$\sqrt{s}=7$, $8$ and $13$ TeV. Section \ref{comparison} is devoted to
a comparison with the currently available CMS data at $\sqrt{s}=8$
TeV. Finally, in Section \ref{conclusions} we draw our conclusions.

%
\section{Leading Order Results with Parton Shower}
\label{framework}
%
%
\begin{figure}
\begin{center}
\includegraphics[width=0.49\textwidth]{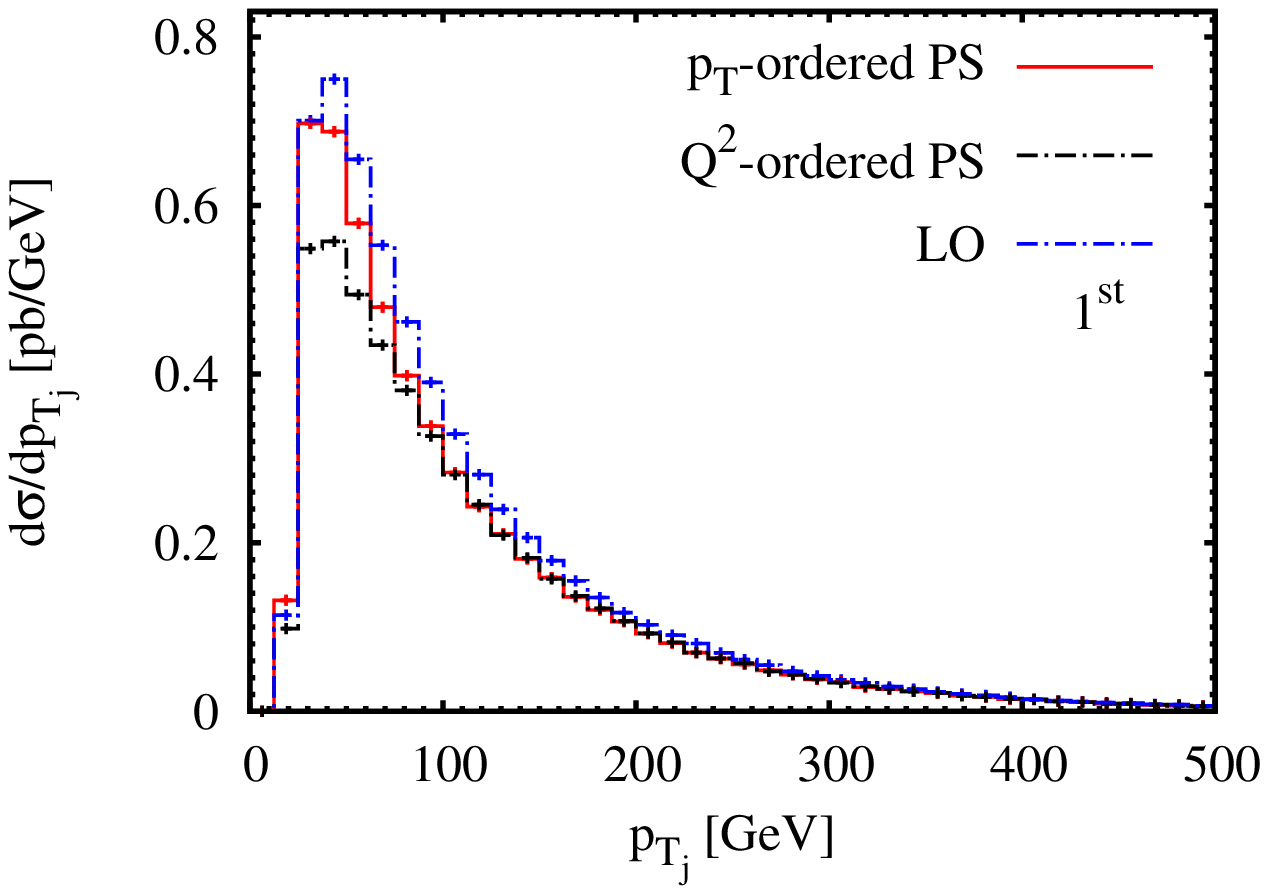}
\includegraphics[width=0.49\textwidth]{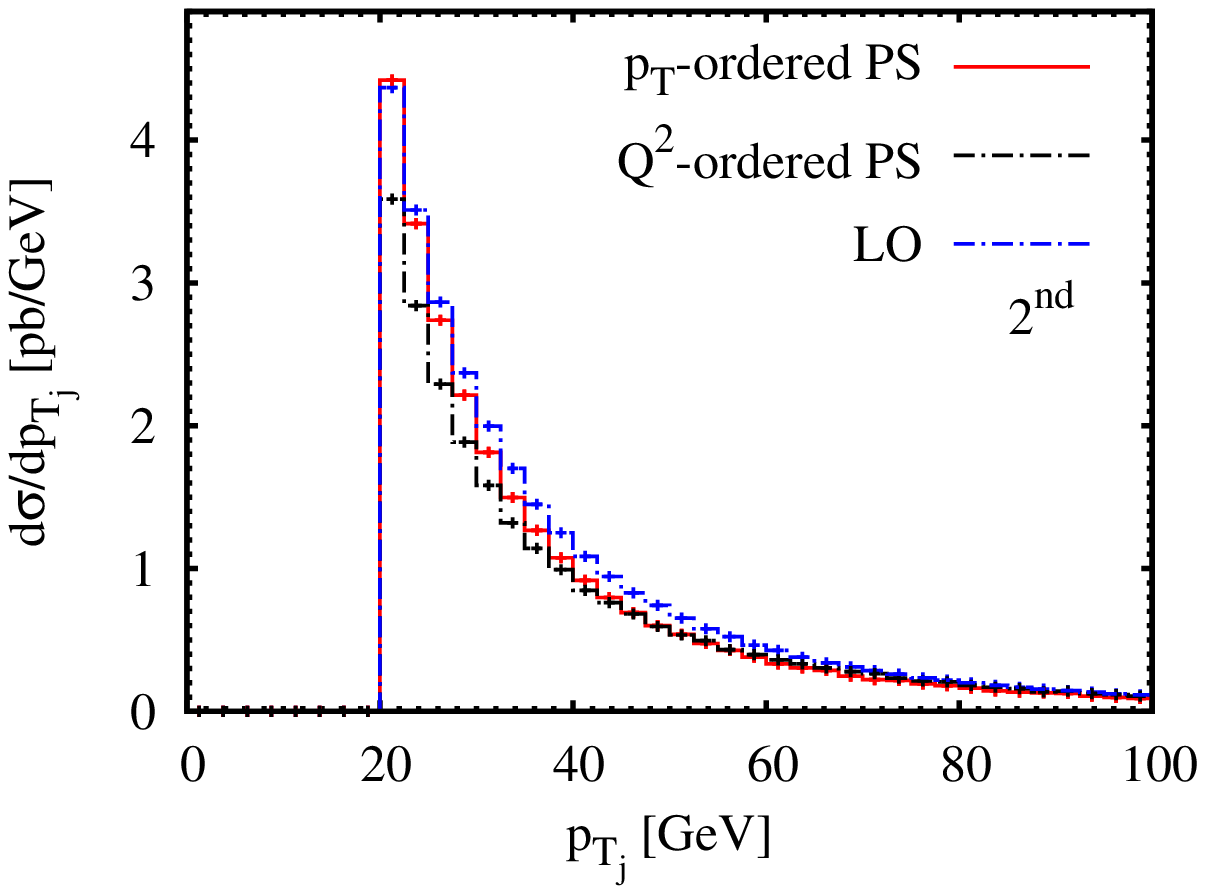}
\includegraphics[width=0.49\textwidth]{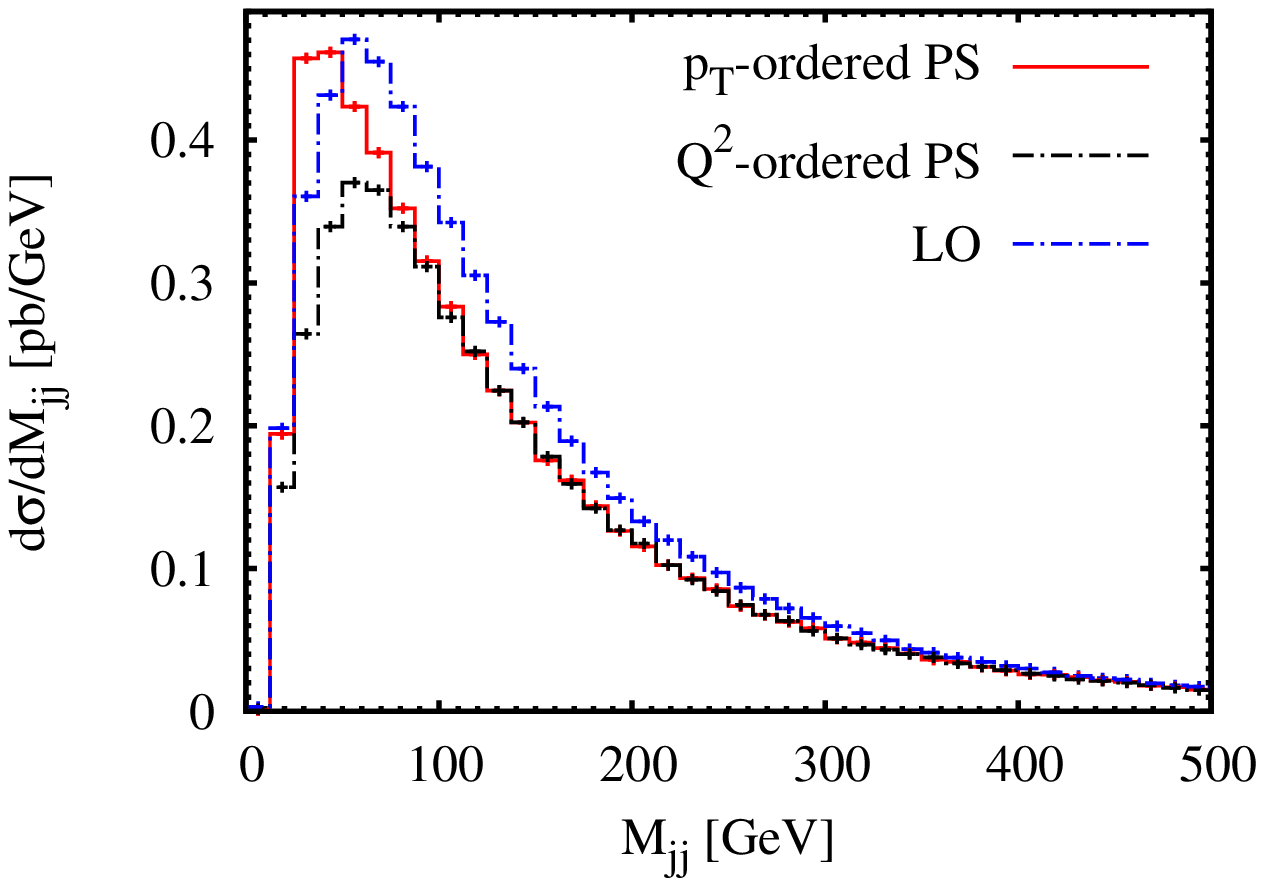}
\includegraphics[width=0.49\textwidth]{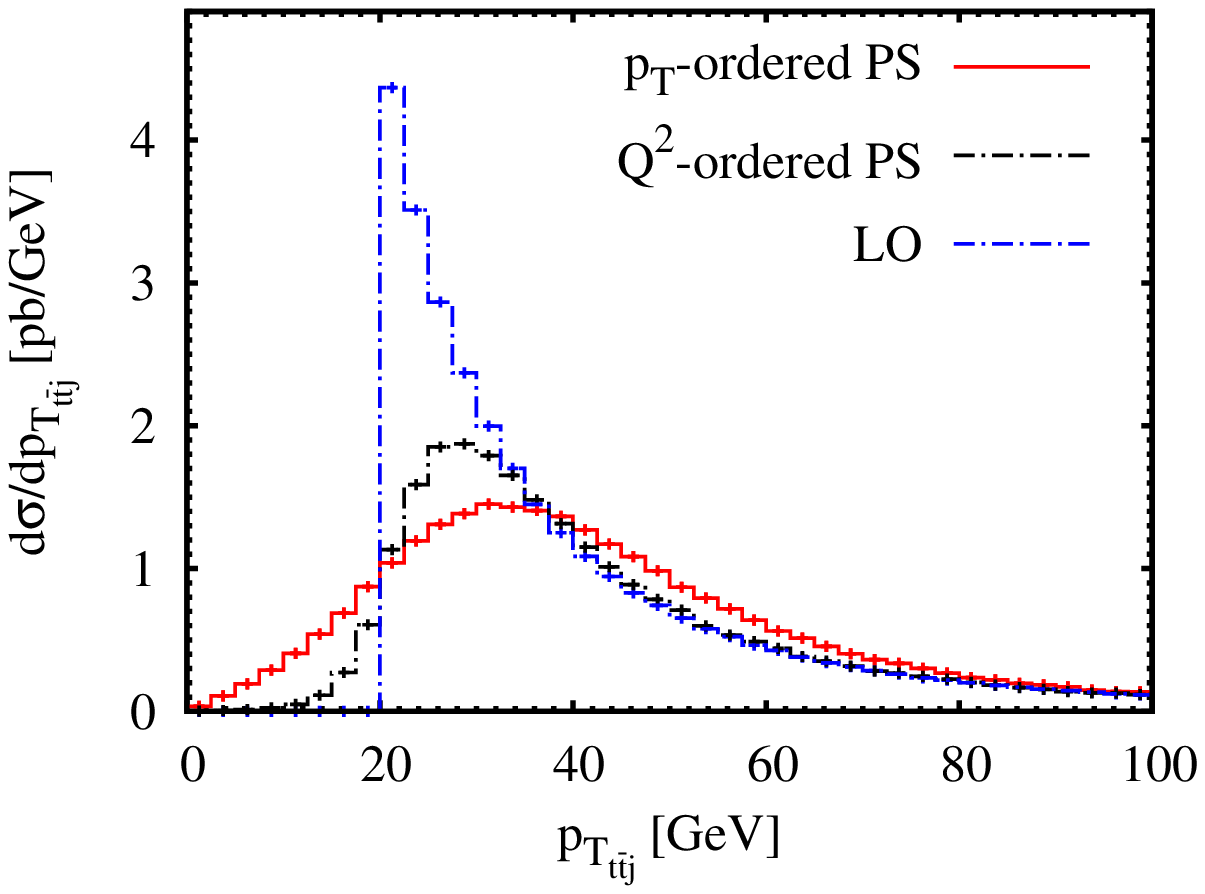}
\end{center}
\caption{\it \label{parton-shower}   Comparison between the LO results
  obtained with  \textsc{Helac-Phegas} and the LO+LL results produced
  by matching LO predictions to \textsc{Pythia} for $pp\rightarrow
  t\bar{t} jj$ at the LHC with $\sqrt{s}$ = 8 TeV. The dash-dotted
  (blue) curve corresponds to the LO whereas the solid (red) to the
  LO+LL based on transverse-momentum-ordered parton shower and the
  dashed (black) curves to the  LO+LL with virtuality-ordered
  shower. The following distributions are shown: transverse momentum
  of the first and the second hardest jet (upper panel), invariant
  mass of the two hardest jets and transverse momentum of the
  $t\bar{t}j$ system (lower panel).}
\end{figure}
%

We begin our analysis by exploring the validity domain of our
perturbative calculation. To this end,  we have
generated an inclusive parton-level sample of $pp\rightarrow
t\bar{t}jj$  where $j$ stands for $u,d,c,s,b$ or $g$. 
The event sample has been
produced with \textsc{Helac-Phegas}
\cite{Kanaki:2000ey,Papadopoulos:2000tt, Cafarella:2007pc} in the Les
Houches event file format \cite{Alwall:2006yp} and interfaced with the
general purpose Monte Carlo program \textsc{Pythia 6.4} (version
6.427) \cite{Sjostrand:2006za} to include initial- and final-state
shower effects. We simulate $pp$ collisions at
$\sqrt{s}=$ 8 TeV using the following parton level cuts, 
\[
p_{T_{j}} =\sqrt{p^2_{x_{j}}+p^2_{y_{j}} }> 10 ~{\rm{GeV}} \,,
\] 
\begin{equation}
|y_j|= \left|\frac{1}{2}\ln \left( \frac{E_j+p_{z_{j}}}{E_j-p_{z_j}}
\right) \right|< 4.5 \,,
\end{equation}
\[
\Delta R_{jj} =\sqrt{\Delta \phi^2_{jj}+\Delta y^2_{jj}}> 0.4 \,,
\]      
where $p_{T_{j}}$, $y_j$ and $\Delta R_{jj}$ denote transverse
momentum, rapidity and distance between the two jets in the $(y,\phi)$
plane respectively.   The top quark mass is set to the
value $m_t$ = 173.5 GeV \cite{Beringer:1900zz} and top quarks are
assumed to be stable.  All the other QCD partons, including bottom
quarks, are treated as massless. We work within the 5-flavor scheme,
taking a LO PDF set with a 1-loop running strong coupling constant and
five active flavors, $N_F=5$.  More specifically, the Les Houches
Accord  PDF implementation
\cite{Giele:2002hx,Whalley:2005nh,Bourilkov:2006cj} of the CT09MC1 PDF
set \cite{Lai:2009ne} is used with the corresponding  value of
$\alpha_s$ evaluated for $\mu=m_{t}$. Jets are reconstructed out of
the partonic final state emerging after shower, using the {\it
anti}-$k_T$ jet clustering algorithm \cite{Cacciari:2008gp} provided
by the \textsc{FastJet} package
\cite{Cacciari:2011ma,Cacciari:2005hq}. The jet cone size is set to
$R=0.5$, and reconstructed jets are required to satisfy
\begin{equation} p_{T_{j}} > 20 ~ {\rm  GeV}, ~~~~~~~~~ |y_j|<2.5, 
~~~~~~~~~~ \Delta R_{jj}> 0.5\,.
\end{equation} 
To allow for a more direct comparison with our fixed-order results, we
decide to stop the evolution at the end of the perturbative phase. In
other words, we neglect effects related to hadronization, underlying
events  or multiple $pp$ interactions. Also, decays of the top quark
and QED radiation from quarks are switched off. All the other
\textsc{Pythia} parameters have been left unchanged and correspond to
default settings. 

We have considered two different variants of shower, both provided
within \textsc{Pythia  6.4}: transverse-momentum ordered shower
(dubbed $\textsc{Pythia}_{p_T}$) and virtuality-ordered or
mass-ordered shower (dubbed $\textsc{Pythia}_{Q^2}$). The starting
scale for the shower has been set to $p^{\rm{min}}_{T_{j}}$ and
$m^{\rm{min}}_{jj}= p^{\rm{min}}_{T_{j}}\sqrt{2(1-\cos R )}$
respectively.  As a consistency check, we have compared the total rate
obtained after showering with the LO expectation based on our
selection cuts. We obtain the following cross sections:
\begin{eqnarray}
 \sigma^{\textsc{Helac+Pythia}_{p_{T}}}_{pp\to t\bar{t}jj}  &=&
69.6 ~{\rm pb} \,,    \nonumber \\ \nonumber  \\
\sigma^{\textsc{Helac+Pythia}_{Q^{2}}}_{pp\to t\bar{t}jj} 
&=& 63.7 ~{\rm pb} \,,
\nonumber \\ 
\nonumber \\
\sigma^{\textsc{Helac-Phegas}}_{pp\to t\bar{t}jj}
&=& 
77.1 ~{\rm pb} \,.
\end{eqnarray}   
The two showered results, based on different shower
ordering variables, agree within $9\%$ and are comparable with the LO
cross section. 

In a subsequent step, we compare leading-order predictions at the
differential level before and after showering. Figure
\ref{parton-shower} shows distributions of the transverse momentum of
the two hardest jets, the dijet invariant mass and the transverse
momentum of the $t\bar{t}j_{1}$ system, where $j_1$ denotes the first
hardest jet. We observe that $p_T$ and invariant mass distributions
are not strongly modified by the parton shower. Shape differences are
within the  corresponding theoretical errors, that we did not report
on the plots for better readability.  On the other hand, the
transverse momentum distribution of the $t\bar{t}j_1$ system shows a
sizeable discrepancy in the low-$p_{T}$ region. Note that at
leading-order, momentum conservation sets the equality
$p_{T}(t\bar{t}j_1) = p_{T}(j_2)$, where $j_2$ is the second hardest
jet, and thus the distributions of these two observables
coincide. When the parton shower is turned on, the extra radiation
allows the presence of additional jets, and the direct relation
between the previous two quantities is lost. A  large Sudakov
suppression is visible starting approximately below $p_{t\bar{t}j_1} =
40$ GeV, while the fixed-order result displays a sharp peak \footnote{
Similar conclusions have been obtained before for example 
in case of $t\bar{t}j$ either by means of  matching  different LO multijet
matrix elements with showering programs \cite{Mangano:2006rw} or
by matching the  NLO $t\bar{t}j$  matrix element with
parton shower via the \textsc{Powheg} method 
\cite{Alioli:2011as}.}.  This discrepancy indicates that dominant
higher-order effects endanger the stability of the perturbative
expansion in the small $p_T$ region for this observable. 
Therefore to be on the safe side for all observables the following choice
of basic selection cuts is taken for a reliable fixed-order 
analysis:
\begin{equation}
\label{final-cuts}
p_{T_{j}} > 40 ~ {\rm GeV}, ~~~~~~~~~ |y_j|<2.5, 
~~~~~~~~~ \Delta R_{jj}> 0.5 \,.
\end{equation}  
The specific value of the cut on the maximum jet
rapidity is dictated by the detector acceptance and the experimental
requirements for the bottom flavor jet reconstruction
\cite{Chatrchyan:2012jua}. We report for completeness the total LO
cross sections that we obtain using the cuts (\ref{final-cuts}):
\begin{eqnarray}
 \sigma^{\textsc{Helac+Pythia}_{p_{T}}}_{pp\to t\bar{t}jj}  &=&
26.4 ~{\rm pb} \,,    \nonumber \\ \nonumber  \\
\sigma^{\textsc{Helac+Pythia}_{Q^{2}}}_{pp\to t\bar{t}jj} 
&=& 23.1 ~{\rm pb} \,,
\nonumber \\ 
\nonumber \\
\sigma^{\textsc{Helac-Phegas}}_{pp\to t\bar{t}jj}
&=& 
28.1 ~{\rm pb} \,.
\end{eqnarray}  

Let us conclude this section by saying that the main point here 
was to justify our choice of the $p_T$ cut on the
jets. We were  not aiming at a very precise description of particular
observables, such  as the $p_T$ of the $t\bar{t}j$ system, over the
complete range of transverse momenta. For this reason, we made some
approximations, which are, in our opinion, justified by our
goal. These  approximations consist of: lack of merging of samples
with different  multiplicity, lack a elimination of double
counting. We stress that a  more involved procedure without these
approximations would not change  our conclusion, which is that a
lowest $p_T$ of $40$ GeV is very safe from  the point of view of the
reliability of the fixed order prediction. To  be very precise, what
mattered to us, was the point where a showered  distribution diverges
from the fixed order one. Merging, on the other  hand, would mostly
improve the low $p_T$ range. 

Let us also stress here, that we have drawn similar conclusions  from
matching  the leading order  $pp\to t\bar{t}b\bar{b}$  event sample
with $p_T$-ordered and $Q^2$-ordered showers from \textsc{Pythia}. In
that case, however, initial state  configurations with a b-quark  have
been neglected in the leading order  $pp\to t\bar{t}b\bar{b}$  event
sample.   Such contributions are usually  neglected in calculations
for final states involving a $b\bar{b}$  pair. The reason is simple:
they contribute at the level of a  percent, as has been checked in
many studies in the past. In view of  the quality of the prediction
such contributions are irrelevant, but  make the technical side of the
work more involved. Of course, a shower  will not change anything
here.

%
\section{Next-to-leading Order Differential Cross Sections}
\label{nlo-distributions}
%
%
\begin{figure}
\begin{center}
\includegraphics[width=0.95\textwidth]{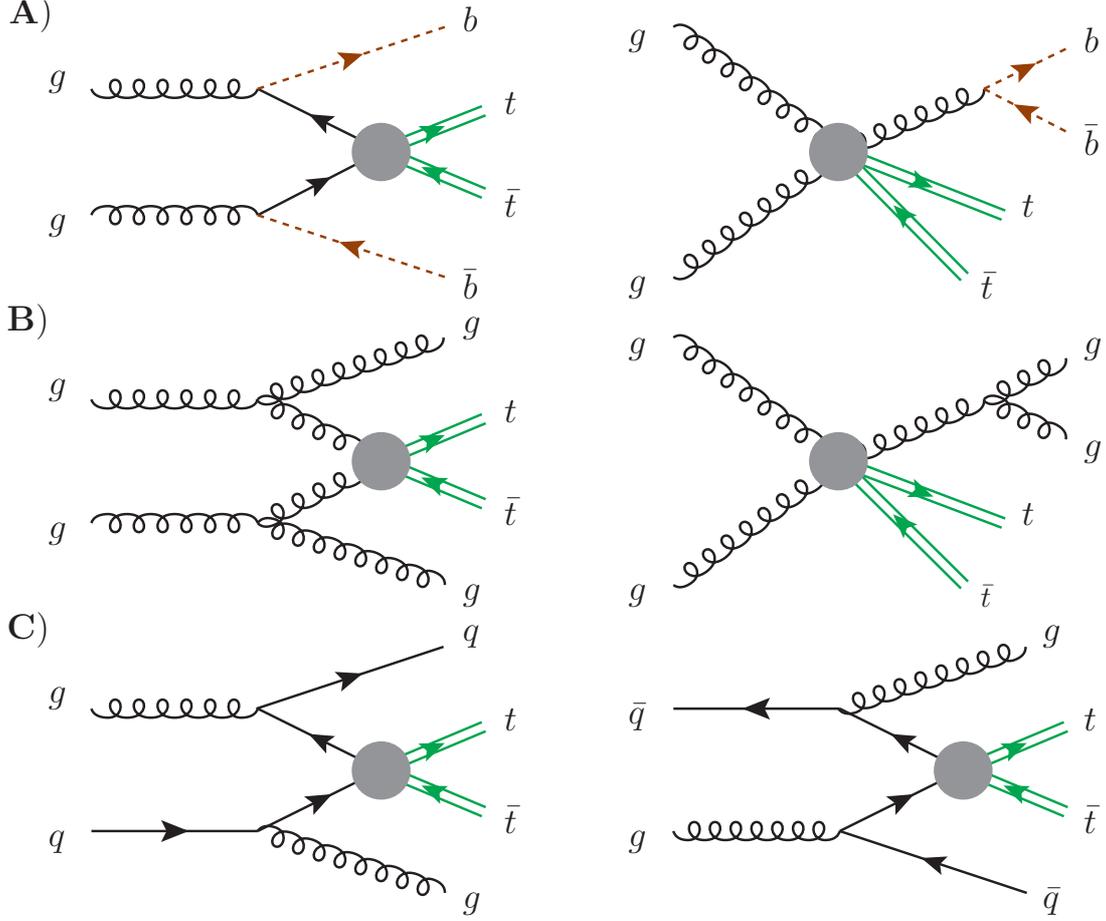}
\end{center}
\caption{\it \label{diagrams}  Feynman diagrams contributing to the
  dominant $gg\rightarrow t\bar{t}b\bar{b}$,  $gg\rightarrow
  t\bar{t}gg$, $gq\rightarrow t\bar{t}gq$ and   $g\bar{q}\rightarrow
  t\bar{t}g\bar{q}$ subprocesses for the following processes
  $pp\rightarrow t\bar{t}b\bar{b}$ and  $pp\rightarrow t\bar{t}jj$
  respectively.  Blobs denote all possible substructures  of the
  corresponding diagram.}  
  \end{figure}
%
\begin{figure}
\begin{center}
\includegraphics[width=0.49\textwidth]{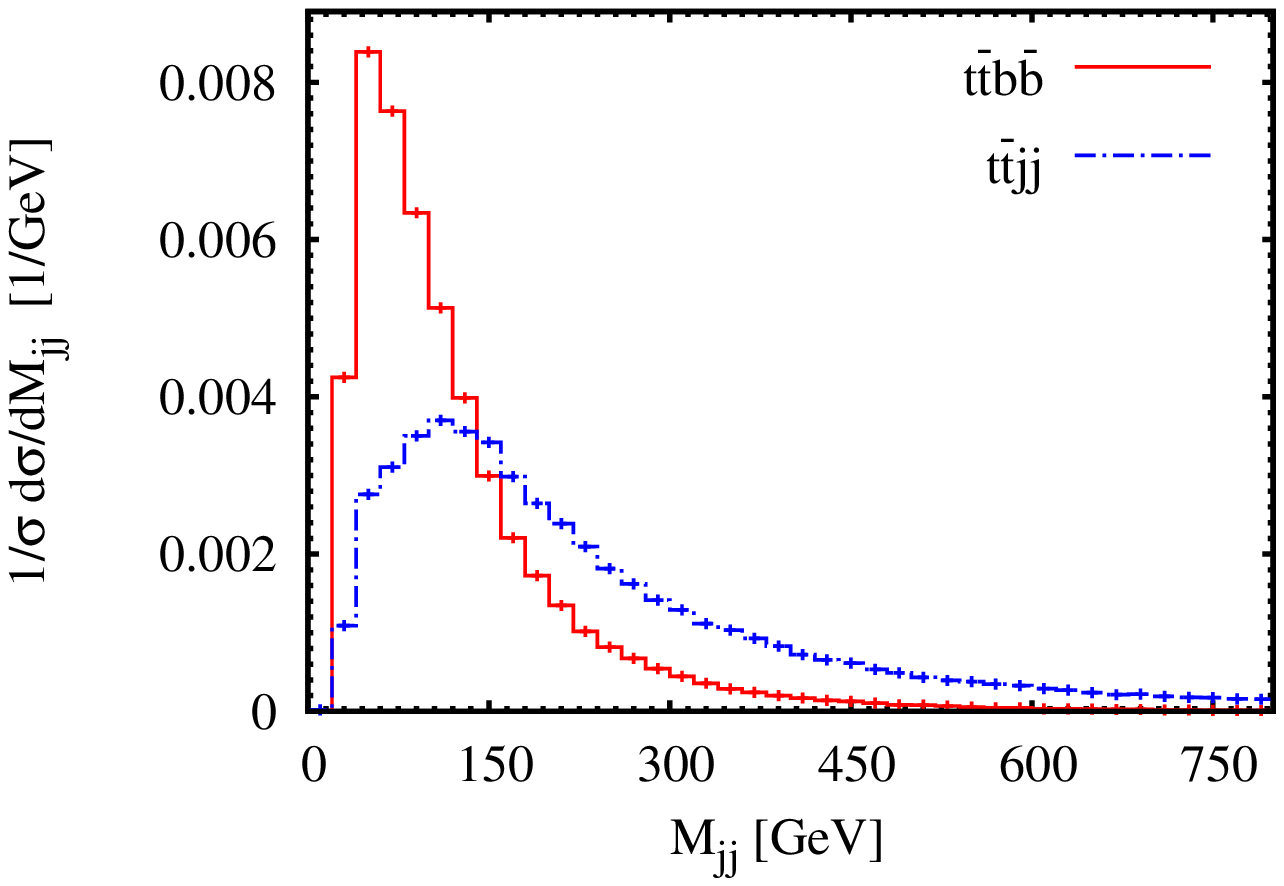}
\includegraphics[width=0.49\textwidth]{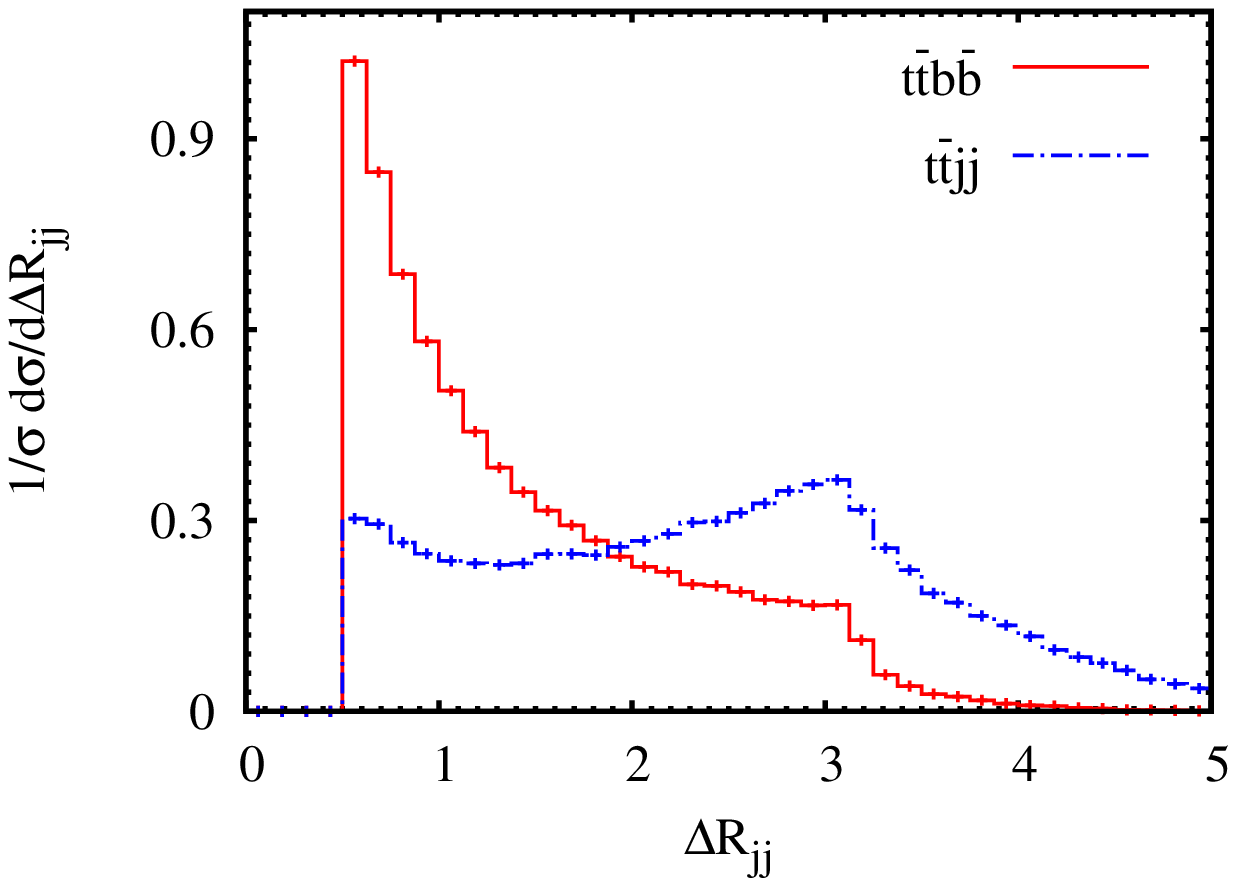}
\includegraphics[width=0.49\textwidth]{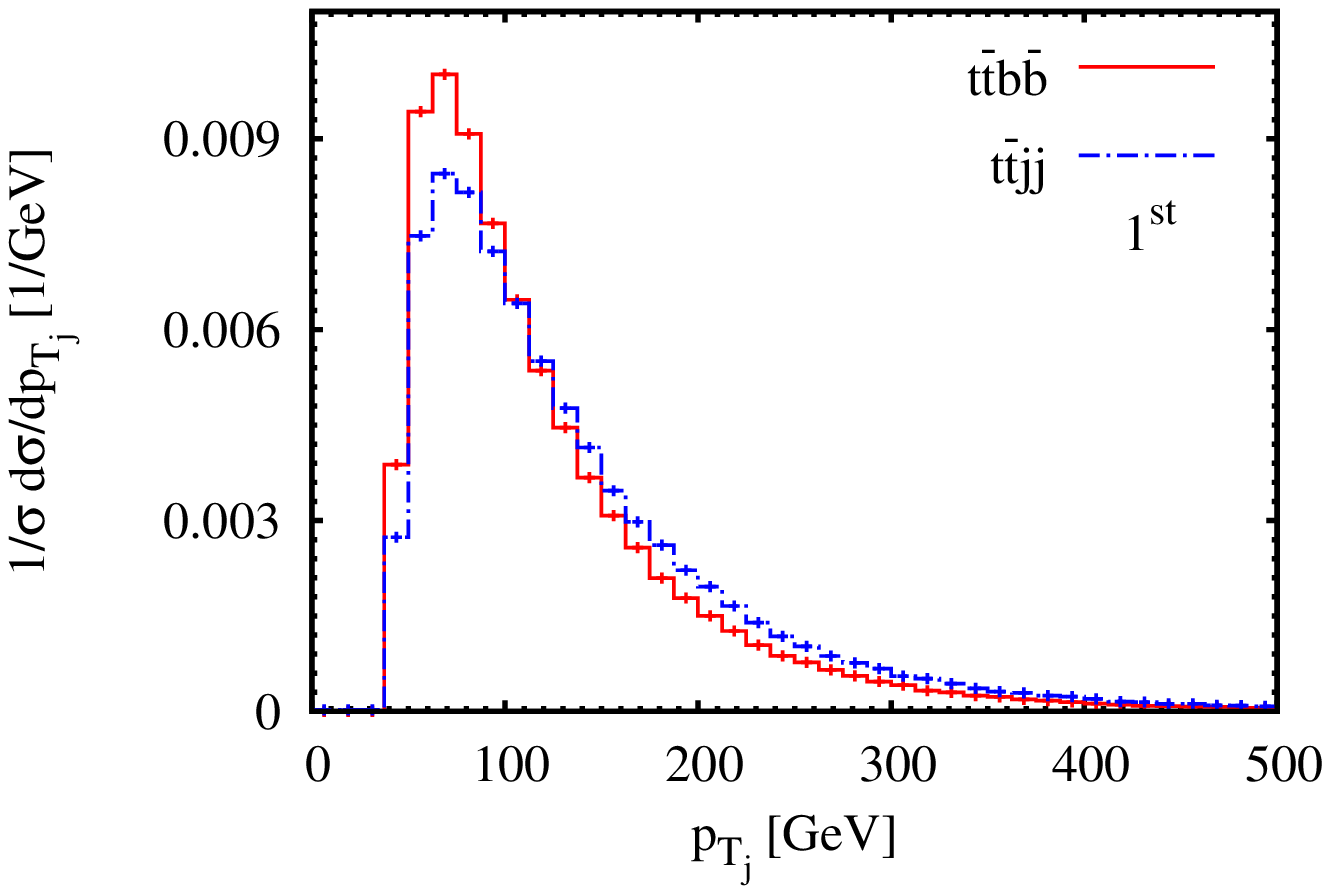}
\includegraphics[width=0.49\textwidth]{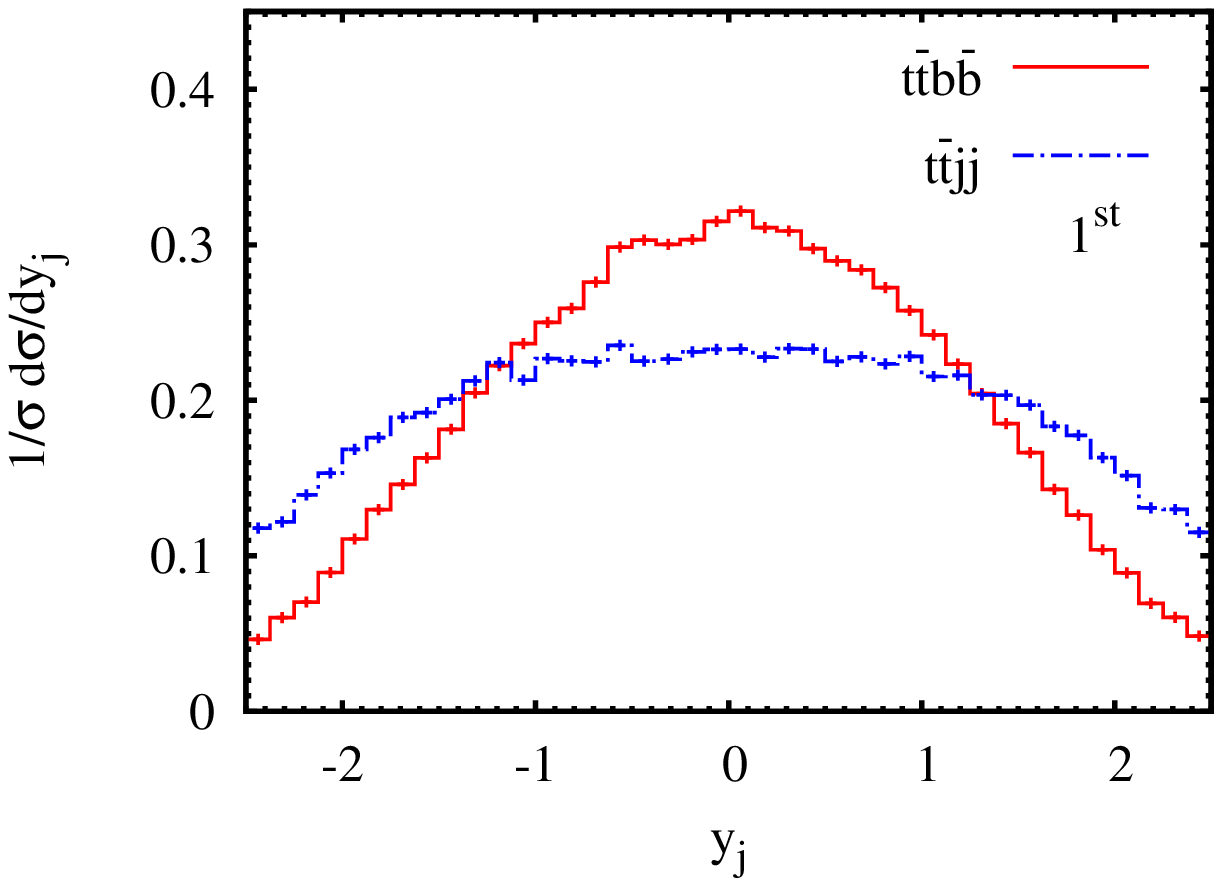}
\includegraphics[width=0.49\textwidth]{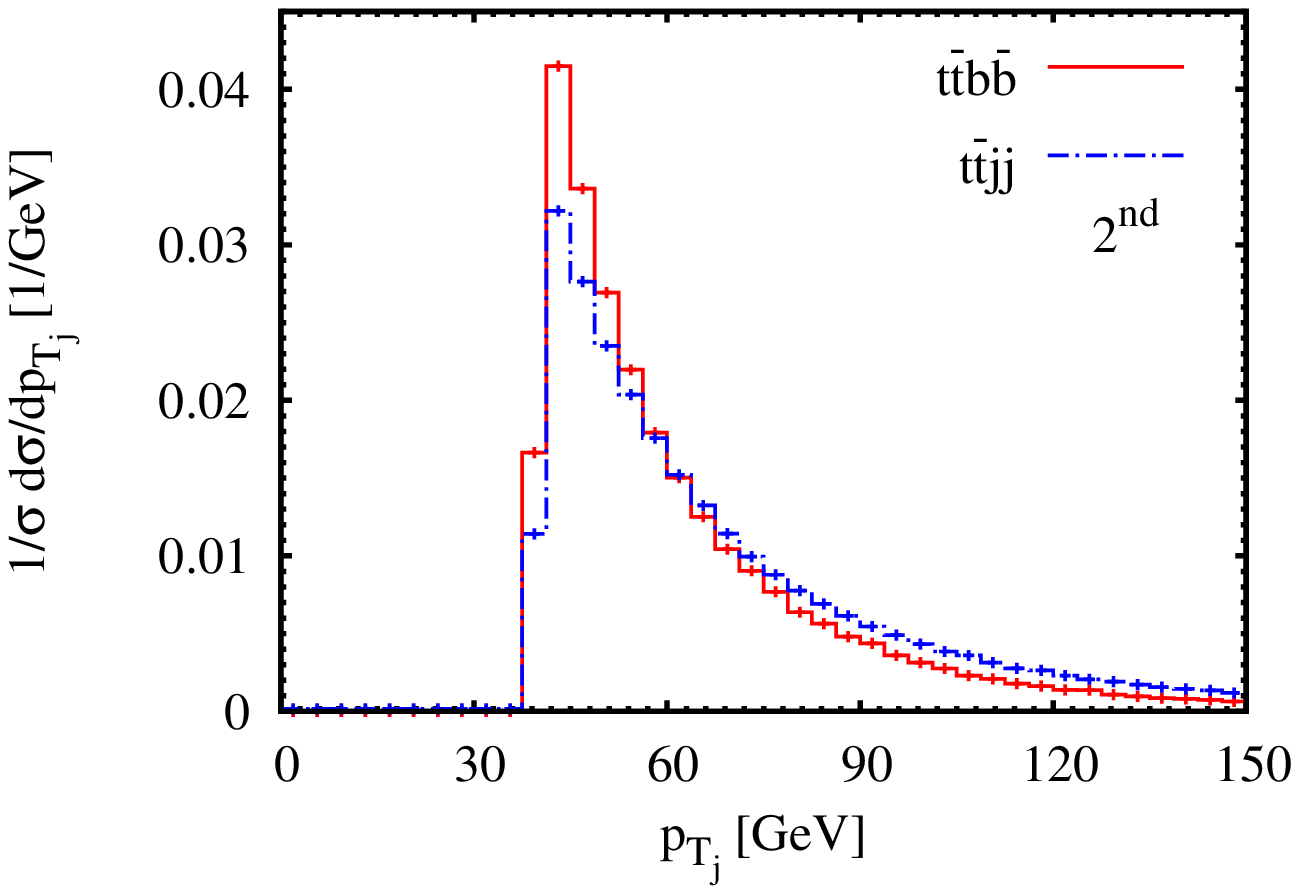}
\includegraphics[width=0.49\textwidth]{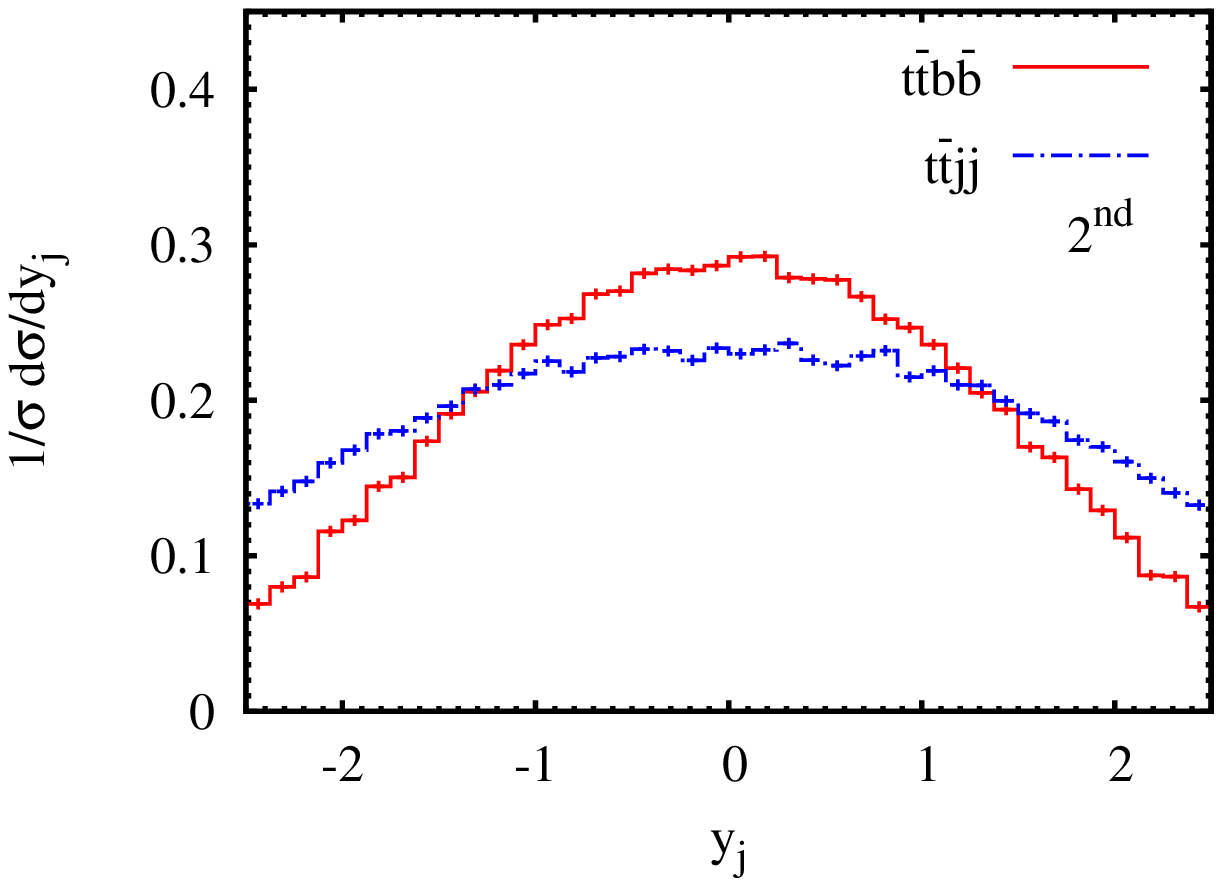}
\end{center}
\caption{\it \label{nlo-jets}  Comparison of the normalized
  next-to-leading order differential  cross sections for
  $pp\rightarrow t\bar{t} b\bar{b}$ and $pp\rightarrow t\bar{t} jj$ at
  the LHC with $\sqrt{s}$ = 8 TeV.  The dash-dotted (blue) curve
  corresponds to the $pp\rightarrow t\bar{t} jj$ process whereas the
  solid (red) curve to $pp\rightarrow t\bar{t} b\bar{b}$.  The
  following distributions are shown: invariant mass of the two hardest
  jets,  separation between those jets (upper panel), transverse
  momentum and rapidity of the first (middle panel) and the second
  hardest jet (lower panel).}  
  \end{figure}
%
\begin{figure}
\begin{center}
\includegraphics[width=0.49\textwidth]{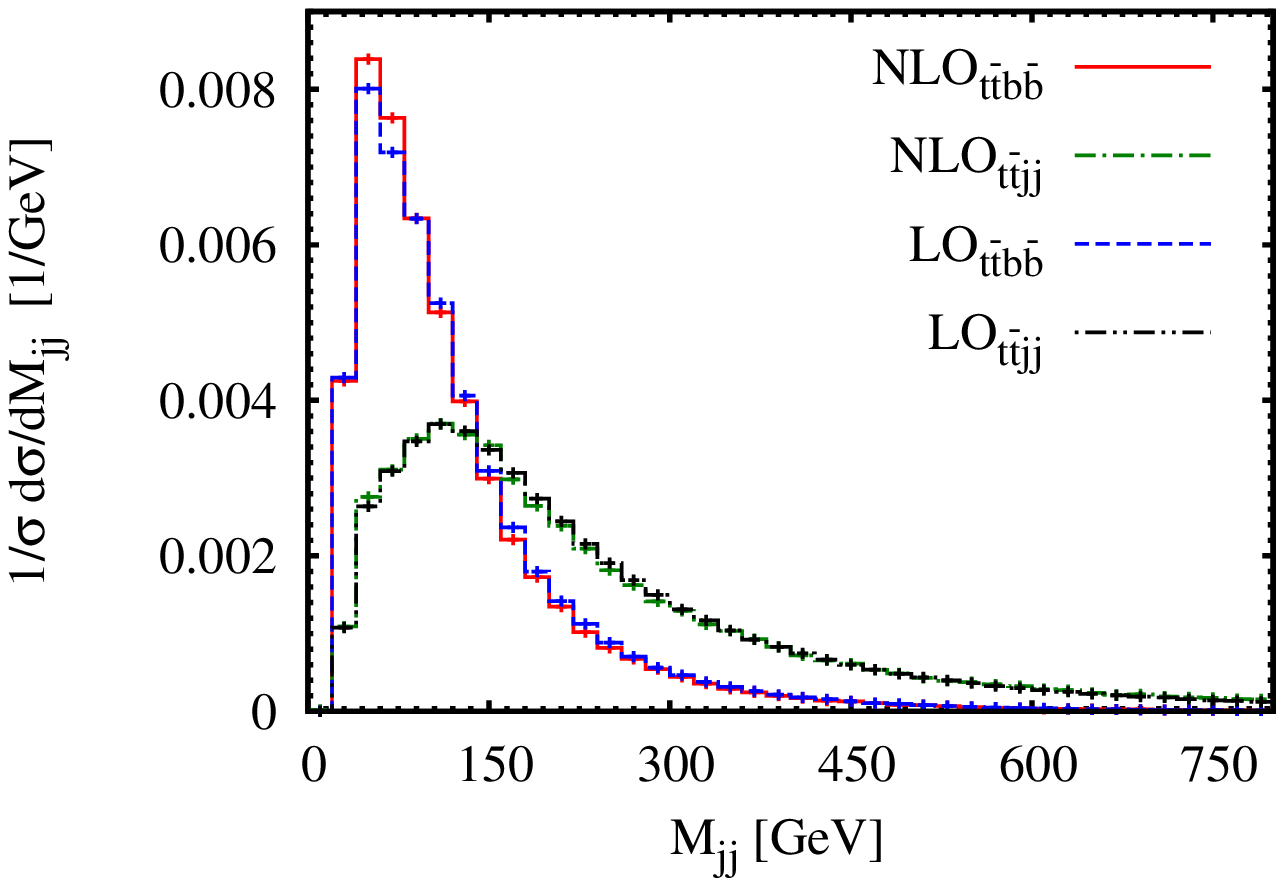}
\includegraphics[width=0.49\textwidth]{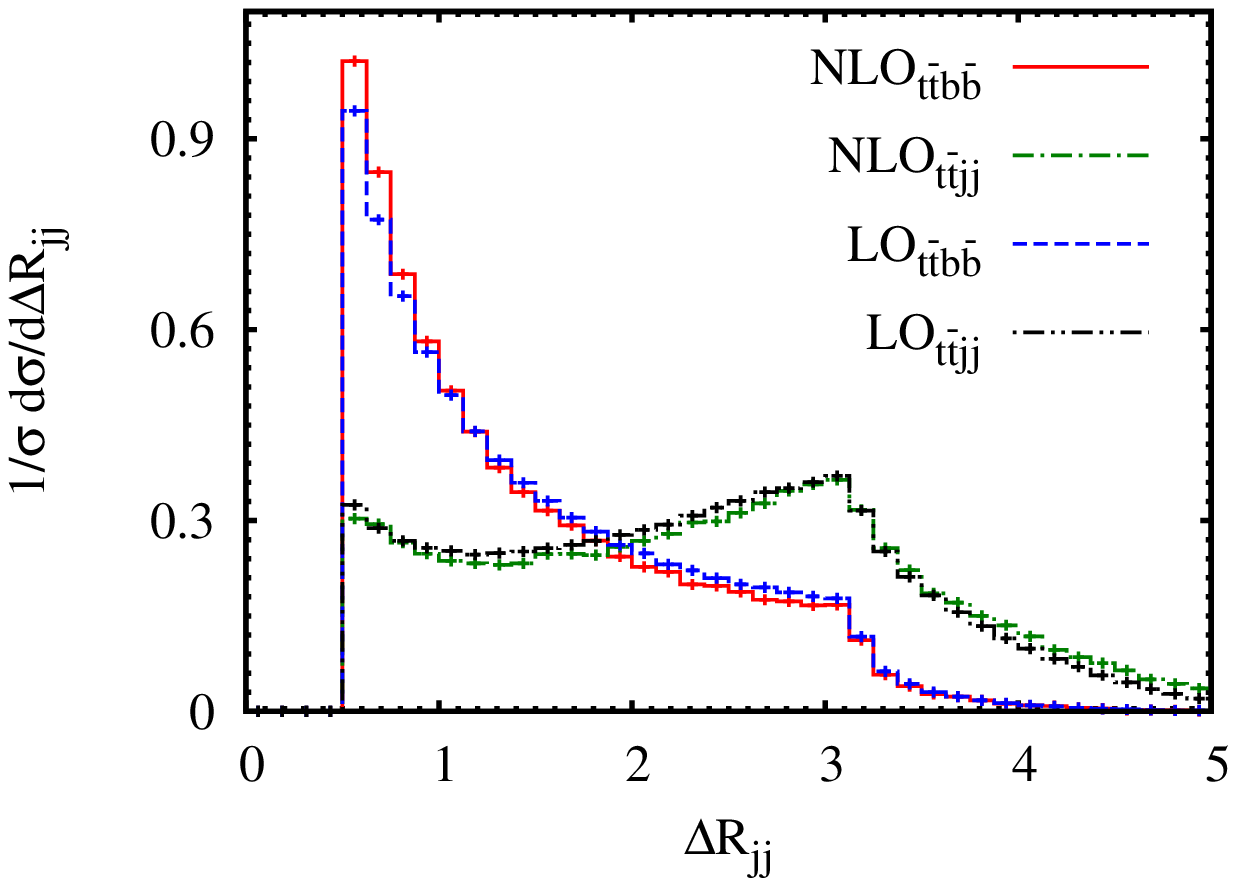}
\includegraphics[width=0.49\textwidth]{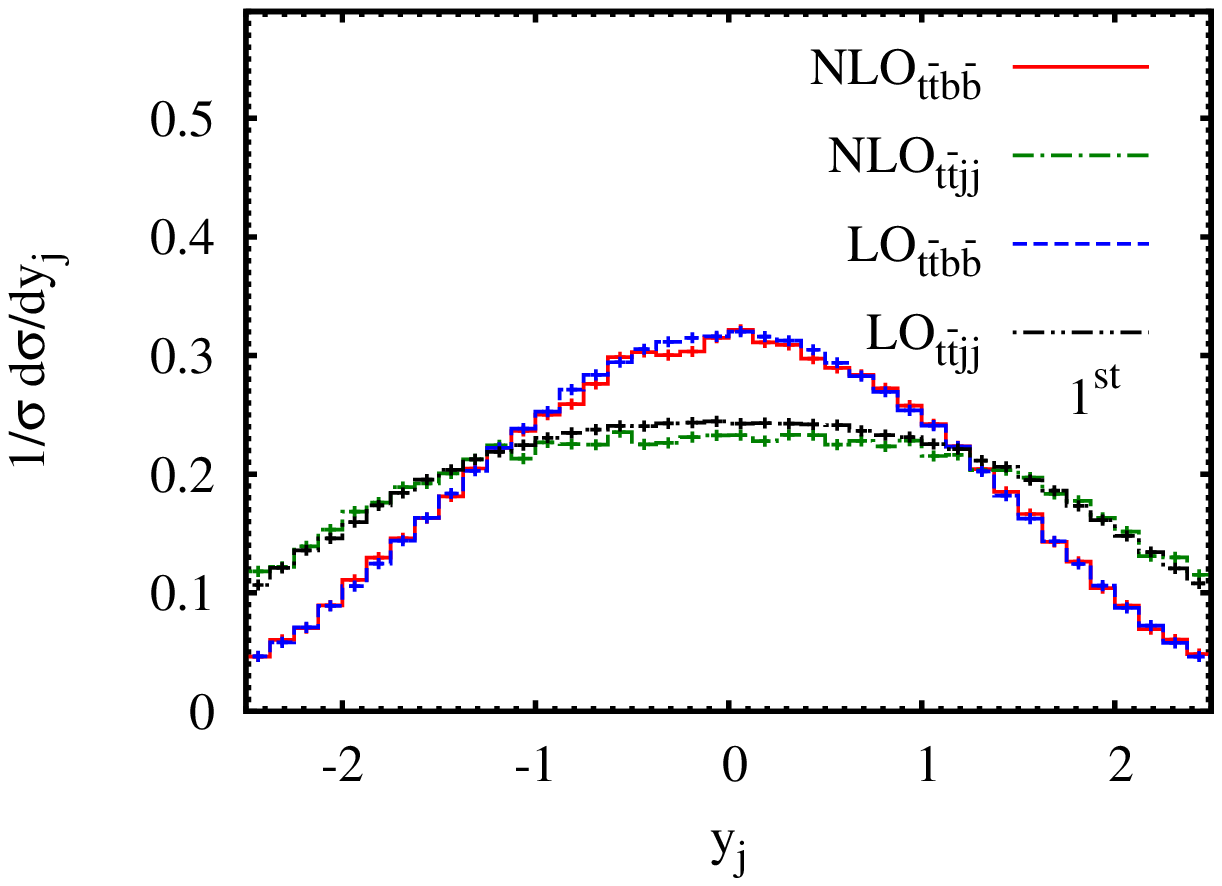}
\includegraphics[width=0.49\textwidth]{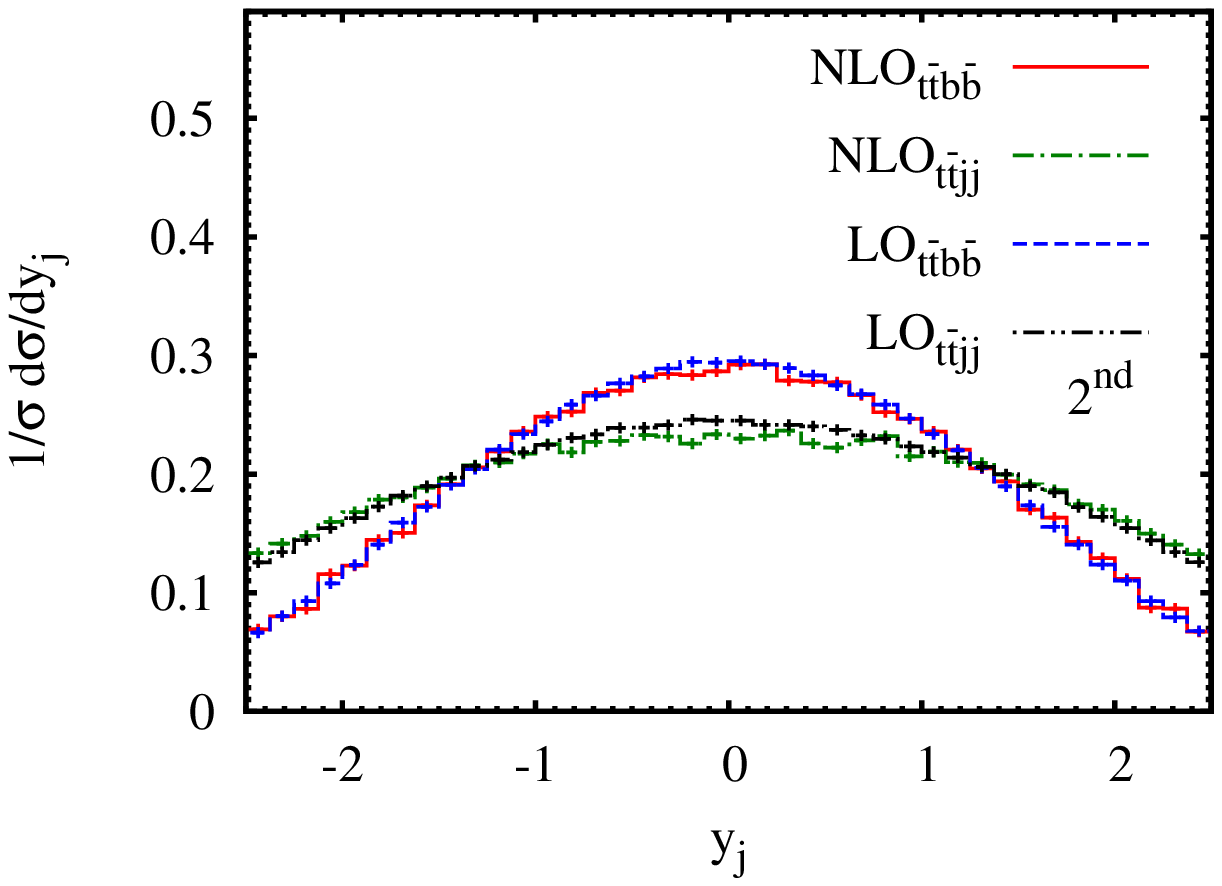}
\end{center}
\caption{\it \label{lo-nlo-jets}  Comparison of the normalized leading order 
  and next-to-leading
  order differential  cross sections for  $pp\rightarrow t\bar{t}
  b\bar{b}$ and $pp\rightarrow t\bar{t} jj$ at the LHC with $\sqrt{s}$
  = 8 TeV.   The following distributions are
  shown: invariant mass of the two hardest jets,  separation between those 
  jets and rapidity of the first and the second hardest  jet.}  
  \end{figure}
%

Having established a safe kinematical domain, we now turn to examine
the behavior of differential cross sections for both processes, $pp\to
t\bar{t}b\bar{b}$ and $pp\to t\bar{t}jj$, where $j$ stands for
$u,d,c,s,b,g$ together with corresponding anti-quarks. Also here
initial state configurations with a b-quark are neglected in
calculations for the $pp\to t\bar{t}b\bar{b}$ final state. As already
mentioned, we are interested in investigating similarities and
correlations between the two backgrounds with the goal of reducing
theoretical uncertainties in the cross section ratio.

Our NLO results are based on NLO CTEQ PDF set, {\it i.e.} CT10
\cite{Lai:2010vv},  including 2 loop running of
$\alpha_s$ and $N_F=5$, with $\mu_R=\mu_F=\mu_0$ for the
renormalization and factorization scales, where \footnote{ This scale
choice for the $pp\to t\bar{t} b\bar{b}$ process has been first
introduced in Ref. \cite{Bredenstein:2010rs}.}
\begin{eqnarray}
\mu_0^2 (pp \to t\bar{t}b\bar{b}) & = & m_t \, \sqrt{\prod_{i=1}^{2}
  p_T(b_i)} ~~\,, \\ \mu_0^2 (pp \to t\bar{t}jj) &
= & m_t^2 \,.
\end{eqnarray}
Jets are reconstructed using the {\it anti}-$k_T$ clustering algorithm
with resolution parameter $R = 0.5$. We require the presence of at
least two jets and impose the selection cuts of
Eq.(\ref{final-cuts}). No restriction on the kinematics of the
possible third jet is applied.

All the next-to-leading order results presented in this paper have
been obtained with the help of the package \textsc{Helac-NLO}
\cite{Bevilacqua:2011xh}, which consists of \textsc{Helac-1loop}
\cite{vanHameren:2009dr,Ossola:2007ax,vanHameren:2010cp} and
\textsc{Helac-Dipoles} \cite{Czakon:2009ss,Bevilacqua:2013iha}. The
integration over the phase space has been achieved using
\textsc{Kaleu} \cite{vanHameren:2010gg}.

To understand similarities and differences between the two
backgrounds, it is helpful to identify the dominant partonic
subprocesses. In the case of $pp \to t\bar{t}b\bar{b}$, at LO in the
perturbative expansion, the most important production mechanism is via
scattering of two gluons (see Figure \ref{diagrams} - A). 
Within our selection cut choice, the $gg$
channel contributes to the total LO cross section by about $90\%$ at
$\sqrt{s} = 8$ TeV. On the other hand, $pp \to t\bar{t}jj$ is governed by 
two equally important channels, namely the $gg$ channel ($49\%$) and the
$qg/gq$ channel ($40\%$) (see Figure \ref{diagrams} - B and C). 
We note that the contribution of the process
$gg \to t\bar{t}q\bar{q}$, which is related to the $t\bar{t}b\bar{b}$
final state amounts to $2.6\%$ and is almost negligible  compared to
the dominant contributions. These facts suggest that the two
backgrounds $t\bar{t}b\bar{b}$ and $t\bar{t}jj$ might show different
features in the jet kinematics. This would have of course a negative
impact on correlations.

A collection of observables is reported in Figure \ref{nlo-jets},
where the NLO distributions have been normalized to the corresponding
absolute cross sections, in order to evidentiate shape differences
between the two processes. We focus here on quantities related to jet
activity, such as rapidity and transverse momentum distributions of
the first and the second hardest jet, invariant mass and separation
between the two jets. Note that the requirement of two hard jets with
a resolution parameter $R = 0.5$ and $p_{T_{j}}^{\rm min}=40$ GeV
implies a lower bound on their invariant mass, of the order of
$m_{jj}^{\rm min} = 19.8$ GeV. 

We observe large shape differences in several observables, in line
with our expectations. First of all, the $b$-jets show a preference
for the central region of the detector in comparison with light
jets. This difference is to be ascribed mainly to the contribution of
the $qg/gq$ channel, which favors the emission of jets at larger
rapidities than the $gg$ channel. Note that, contrary to the
$t\bar{t}jj$ case, in $t\bar{t}b\bar{b}$ production the $qg/gq$
channel is absent at LO and becomes available only at NLO.

In general, jets from the $t\bar{t}jj$ background show a much harder
spectrum compared to $t\bar{t}b\bar{b}$. Sizeable differences can also be
seen in the invariant mass and $\Delta R_{jj}$ separation between the
two jets. In fact, using our cut selection, the $t\bar{t}b\bar{b}$
background is  dominated by the $gg\to t\bar{t}g (g \to b\bar{b})$
production mechanism (see Figure \ref{diagrams} - A.2), which naturally
favors the production of $b$-jet pairs with small invariant mass. In
the case of $t\bar{t}jj$,  there is an interplay between two different
mechanisms. On the one hand, $gg\to t\bar{t}g (g \to gg)$ (Figure
\ref{diagrams} - B.2) is relevant for small values of $m_{jj}$ and
gives a  signature quite similar to the $b\bar{b}$ case. On the other
hand, gluon radiation off initial-state partons  (see {\it
  e.g} Figure \ref{diagrams} - B.1) provides an equally
important contribution due to collinear enhancements. Thus, light jets
with large rapidities and large $\Delta R_{jj}$ separation are also likely
to be produced in the $t\bar{t}jj$ case, which explains the quite
different $\Delta R_{jj}$ spectrum.  All the kinematical features described
above are rather insensitive to higher-order corrections as shown
in Figure \ref{lo-nlo-jets}, where we compare  normalized LO and NLO
differential cross sections.

\begin{figure}[t!]
\begin{center}
\includegraphics[width=0.49\textwidth]{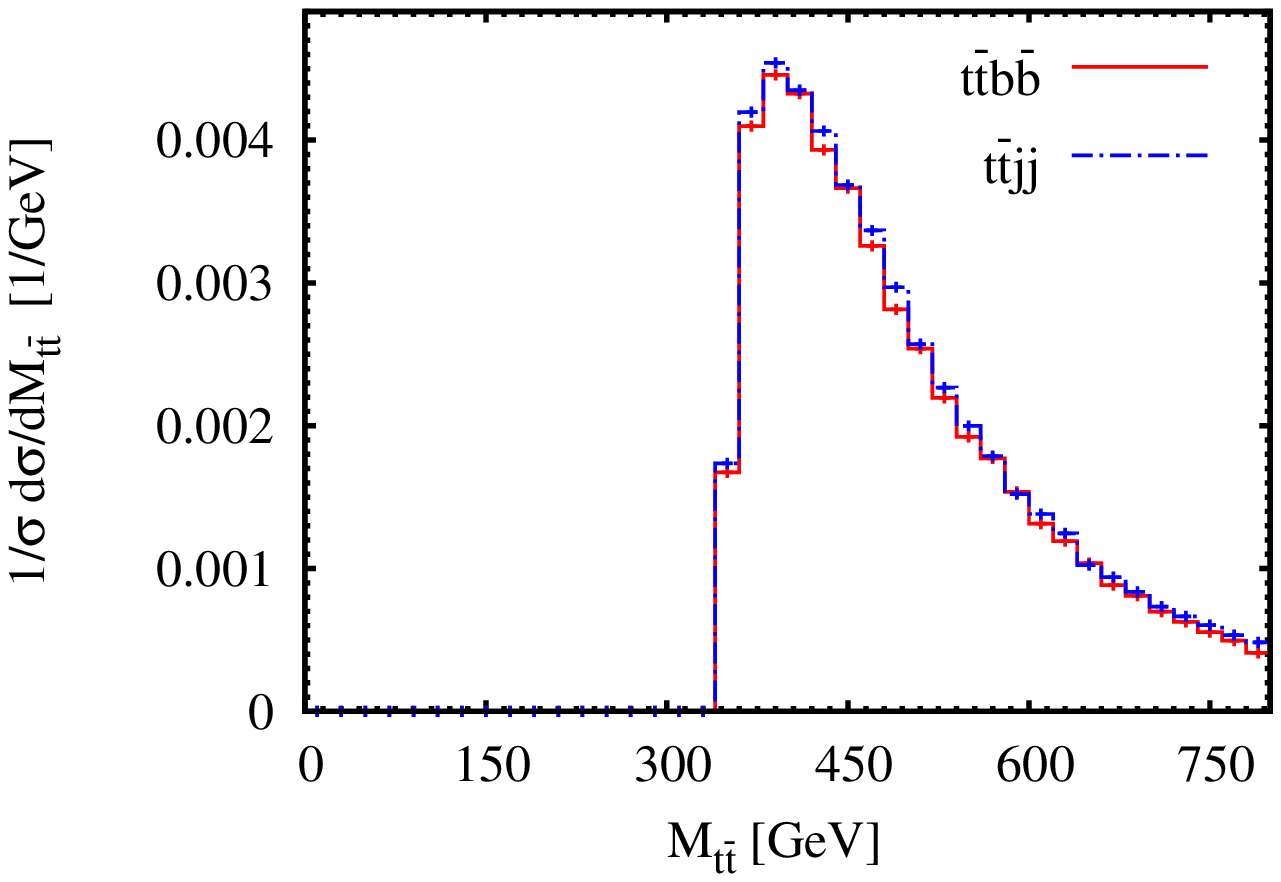}
\includegraphics[width=0.49\textwidth]{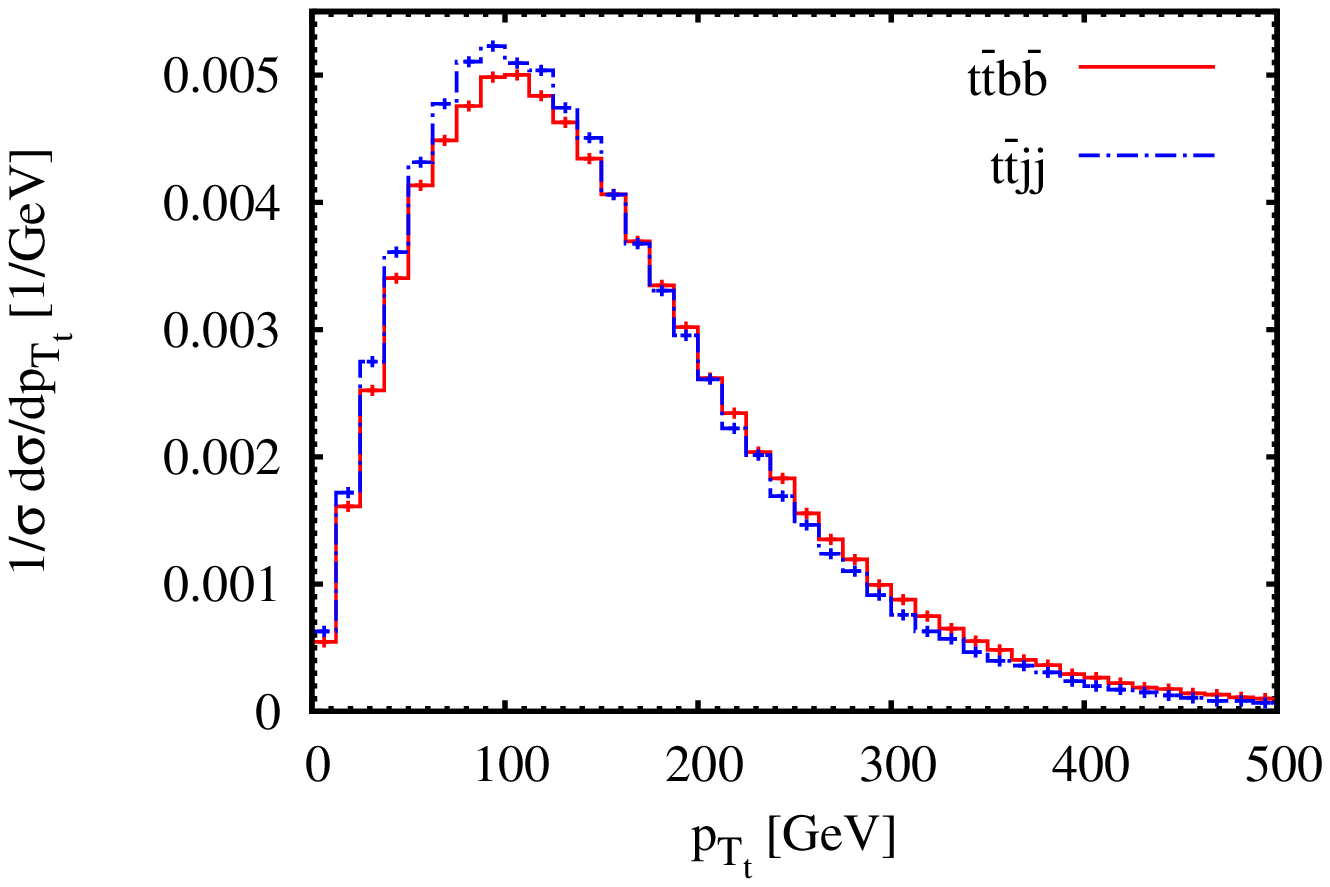}
\end{center}
\caption{\it \label{nlo-tops}   Comparison of the normalized next-to-leading
  order differential  cross sections for  $pp\rightarrow t\bar{t}
  b\bar{b}$ and $pp\rightarrow t\bar{t} jj$ at the LHC with $\sqrt{s}$
  = 8 TeV.  The dash-dotted (blue) curve corresponds to the
  $pp\rightarrow t\bar{t} jj$ process whereas the solid (red) curve to
  $pp\rightarrow t\bar{t} b\bar{b}$.  The following distributions are
  shown: invariant mass of the $t\bar{t}$ system and (averaged) 
  transverse  momentum of the top quark.}
  \end{figure}

Despite sizeable differences in the jet activity, it might still be
possible that $t\bar{t}b\bar{b}$ and $t\bar{t}jj$ show some similarity
connected to the underlying basic process they have in common, {\it
  i.e.} top quark pair production. To this end, we report in Figure
\ref{nlo-tops} normalized distributions of a few observables related
to the top quark kinematics, namely invariant mass of the $t\bar{t}$
system and averaged transverse  momentum of top quarks. Indeed,
distributions show a very good agreement in shape, indicating some
level of correlation. The pretty different jet kinematics that
characterizes the two backgrounds has a minimal influence on the
underlying heavy $t\bar{t}$ system.

 
%
\section{Next-to-leading Order Cross Section Ratio}
\label{nlo-ratio}
%

In this Section we present NLO predictions for the ratio
$\sigma_{t\bar{t}b\bar{b}}/\sigma_{t\bar{t} jj}$ at the LHC for
$\sqrt{s}=$ 7, 8 and 13 TeV. In addition to the basic selection cuts
of Eq.(\ref{final-cuts}), we also report results for $R = 0.8$ and
$\Delta R_{jj} > 0.8$ to check whether the impact of higher-order
corrections is stable against these two parameters. Indeed we want to be
confident that our choice $\Delta R^{\rm min}_{jj} = R=  0.5$ is well
within the range of stability of the perturbative expansion.

%
\subsection{LHC $@$ 7 TeV}
%
%
We start with the LHC results at $\sqrt{s}=$ 7 TeV. In Table
\ref{tab_7TeV}, absolute NLO cross sections are shown together with
their ratio for two different values of the jet resolution parameter
$R$ and jet separation cut $\Delta R_{jj}$. 
%
\begin{table}[h!]
\vspace{0.4cm}
\begin{center}
  \begin{tabular}{c|c|c|c}
&&&\\
$\sqrt{s}=$ 7 TeV & $\sigma^{\rm NLO}_{pp\to t\bar{t}b\bar{b} }$ [fb] &
$\sigma^{\rm NLO}_{pp\to t\bar{t} jj }$ [pb]        &
\textsc{Ratio} \\
&&&\\
\hline \hline
&&&\\
$\Delta R_{jj} > 0.8$, $R=0.8$& 119.2 $\pm$ 0.1  
& 13.66 $\pm$ 0.02 & 0.0087\\
&&&\\
\hline 
&&&\\
$\Delta R_{jj}> 0.5$, $R=0.5$ & 142.2 $\pm$ 0.2 
& 13.55  $\pm$ 0.02 & 0.0105\\
&&&\\
  \end{tabular}
\end{center}
\caption{\label{tab_7TeV}\it NLO cross sections for $pp \to
  t\bar{t}b\bar{b} $ and  $pp \to t\bar{t} jj$ at the LHC with $\sqrt{s}$ = 7
  TeV, including errors from the Monte Carlo integration. The ratio
  $\sigma^{\rm NLO}_{pp\to t\bar{t}b\bar{b}}/\sigma^{\rm NLO}_{pp\to
    t\bar{t} jj }$ is  also shown. Results for two different values of
  jet resolution parameter $R$ and jets separation cut $\Delta R_{jj}$
  are reported.}
\end{table}
%
We observe that the cross section ratio is rather sensitive to the
variation of those parameters. Decreasing $\Delta R_{jj}$ and $R$ from
$0.8$ to $0.5$ results in $+21\%$ change that is mostly due to a
large, $+19\%$ shift in the NLO $t\bar{t}b\bar{b}$ cross section. The
NLO $t\bar{t}jj$ cross section on the other hand is affected only by
$-1\%$.  Since at the NLO jets may have some structure, {\it i.e.} two
partons can be inside a jet, an interplay between two different
effects can be observed. On the one hand, the simultaneous decrease of the
$\Delta R_{jj}$ separation cut results in higher total NLO cross
sections.  On the other hand, a smaller resolution parameter $R$ means
that the probability of parton radiation outside the area with
distance $R$ is higher. This may be translated into a larger number of
soft jets with $p_{T_{j}} < p_{T_{j}}^{\rm min}$ and lower total NLO
cross section.  Since for the $t\bar{t}jj$ final state  many
events are concentrated around $\Delta R_{jj} = \pi$,  the NLO cross
section is mildly affected by a change in $\Delta R_{jj}$ cut from 0.8
to 0.5.  Accordingly, the effect associated with the resolution
parameter $R$ dominates leading to the lower NLO cross section.  

With $\Delta R_{jj}>0.5$ and $R=0.5$, {\it i.e.} for the values that have been
used in the experimental studies \cite{CMS4}, our predictions for the
absolute cross sections read 
\begin{equation}
\sigma^{\rm NLO}_{t\bar{t}b\bar{b}}({\rm LHC_{7 TeV}},m_t=173.5 ~{\rm
  GeV,CT10})=  142.2^{+24.1(17\%)}_{-34.6(24\%)} ~{\rm [fb]}\,,
\end{equation}
and
\begin{equation}
\sigma^{\rm NLO}_{t\bar{t}jj}({\rm LHC_{7 TeV}},m_t=173.5 ~{\rm
  GeV,CT10})=  13.55^{-1.66(14\%)}_{-1.92(14\%)}  ~{\rm [pb]}\,.
\end{equation}
The theoretical uncertainty associated with neglected higher-order
terms in the perturbative expansion, can be estimated by varying the
renormalization and factorization scales up and down by a factor 2
around the central scale of the process, {\it i.e.} $\mu_0$. The scale
dependence is indicated by the upper/lower value, which corresponds to
$0.5\mu_0$/$2\mu_0$. Our estimated scale uncertainties for the
integrated cross sections  are of the order $14\%-24\%$ ($14\%-20\%$
after symmetrisation).  In addition, we find that the size of the NLO
QCD corrections is moderately affected by lowering both $\Delta
R_{jj}$ and $R$, {\it i.e.} changes of the order of $15\%$ or less are
visible. Since those changes are within our theoretical errors, we
conclude that $\Delta R^{\rm min}_{jj} = R=0.5$ is  still
perturbatively valid and a fixed-order NLO calculation can be
considered reliable.

We now turn to estimating the theoretical error for the cross section
ratio. Given that there is no unique prescription for this in the
literature, we decided to evaluate it using three different
approaches. The first one assumes that the two background processes
are not correlated, and consists in calculating all possible cross
section ratios:
$R=t\bar{t}b\bar{b}(\mu_1)/t\bar{t}jj(\mu_2)$,  where $\mu_1,\mu_2 \in
(0.5\mu_0,\mu_0,2\mu_0)$. All possible combinations are considered,
namely $(\mu_1,\mu_2)= \{ (2,2)$, $(2,1)$, $(2,0.5)$, $(0.5,2)$,
$(0.5,1)$, $(0.5,0.5)$,  $(1,0.5)$ and $(1,2) \}$. The theoretical
error band is determined taking the minimum and maximum values of the
resulting ratios. This approach, that we name {\it uncorrelated},
gives the following result: 
\begin{equation}
\sigma^{\rm NLO}_{t\bar{t}b\bar{b}}/\sigma^{\rm NLO}_{t\bar{t}jj}
({\rm uncorrelated}) = 0.0105^{+0.0038(36\%)}_{-0.0026(25\%)} \,.
\end{equation}
After symmetrisation of the error estimate, we get a scale uncertainty
of $30\%$ for the cross section ratio.

The second approach assumes that some degree of correlation exists, so
the possible combinations to be evaluated are restricted to the subset
$(\mu_1,\mu_2)= \{ (2,2)$, and $(0.5,0.5) \}$. If $t\bar{t}b\bar{b}$
and $t\bar{t}jj$ are indeed correlated, a reduction of the scale
uncertainty in the ratio should be expected. Using this approach,
named {\it correlated}, we get the following result:
\begin{equation}
\sigma^{\rm NLO}_{t\bar{t}b\bar{b}}/\sigma^{\rm NLO}_{t\bar{t}jj}
({\rm correlated}) = 0.0105^{+0.0034(32\%)}_{-0.0013(12\%)} \,.
\end{equation}
Only a minor reduction in the size of the scale uncertainty is
observed. The theoretical error band for the ratio is now $22\%$ and
is of the same order as the error for the absolute cross sections.

The third and last approach uses the relative errors of the absolute
cross sections as input. We assume these quantities as uncorrelated
and add the errors in quadrature, separately for the cases $0.5\mu_0$
and $2\mu_0$. This approach, that we name {\it relative error}, gives
the result
\begin{equation}
\sigma^{\rm NLO}_{t\bar{t}b\bar{b}}/\sigma^{\rm NLO}_{t\bar{t}jj}
({\rm relative ~error}) = 0.0105^{+0.0022(21\%)}_{-0.0029(28\%)}  \,.
\end{equation}
After symmetrisation of the error estimate, the final scale
uncertainty is $24\%$.

%
\subsection{LHC $@$ 8 TeV}
%

\begin{table}[h!]
\vspace{0.4cm}
\begin{center}
  \begin{tabular}{c|c|c|c}
&&&\\
$\sqrt{s}=$ 8 TeV & $\sigma^{\rm NLO}_{pp\to t\bar{t}b\bar{b}}$ [fb] &
$\sigma^{\rm NLO}_{pp\to t\bar{t} jj}$     [pb]   &
\textsc{Ratio} \\
&&&\\
\hline \hline
&&&\\
$\Delta R_{jj} > 0.8$, $R=0.8$& 190.7 $\pm$ 0.2 
& 21.15 $\pm$ 0.02& 0.0090\\
&&&\\
\hline 
&&&\\
$\Delta R_{jj} > 0.5$, $R=0.5$ &229.3  $\pm$   0.3 
& 20.97 $\pm$ 0.03 &0.0109\\
&&&\\
  \end{tabular}
\end{center}
\caption{\label{tab_8TeV} \it  NLO cross sections for $pp \to
  t\bar{t}b\bar{b} $ and  $pp \to t\bar{t} jj$ at the LHC with $\sqrt{s}$ = 8
  TeV, including errors from the Monte Carlo integration. The ratio
  $\sigma^{\rm NLO}_{pp\to t\bar{t}b\bar{b}}/\sigma^{\rm NLO}_{pp\to
    t\bar{t} jj }$ is  also shown. Results for two different values of
  jet resolution parameter $R$ and jets separation cut $\Delta R_{jj}$
  are reported.}
\end{table}

We repeat the same procedure for the case $\sqrt{s}=$ 8 TeV. The NLO
cross sections are reported in Table \ref{tab_8TeV}, together with the
cross section ratio for the two different jet separation cuts and jet
resolution parameters. Our conclusions are similar to the case of
$\sqrt{s} =$ 7 TeV and therefore will be briefly summarized here. The
absolute cross sections and corresponding theoretical errors for
$\Delta R^{\rm min}_{jj}=R=0.5$ are:
\begin{equation}
\sigma^{\rm NLO}_{t\bar{t}b\bar{b}}({\rm LHC_{8 TeV}},m_t=173.5 
~{\rm GeV,CT10})= 229.3^{+40.7(18\%)}_{-55.7(24\%)}  ~{\rm [fb]}\,,
\end{equation}
\begin{equation}
\sigma^{\rm NLO}_{t\bar{t}jj}({\rm LHC_{8 TeV}},m_t=173.5 
~{\rm GeV,CT10}) = 20.97^{-3.25(15\%)}_{-2.79(13\%)} ~{\rm [pb]}\,.
\end{equation}
Accordingly, results for the cross section ratio are presented, and
scale uncertainties evaluated according to the three methods described
in the previous Subsection:
\begin{eqnarray}
 \sigma^{\rm NLO}_{t\bar{t}b\bar{b}}/\sigma^{\rm NLO}_{t\bar{t} jj}
({\rm uncorrelated}) &=& 0.0109^{+0.0043(39\%)}_{-0.0026(24\%)} \,,
     \nonumber \\ \nonumber  \\
\sigma^{\rm NLO}_{t\bar{t}b\bar{b}}/\sigma^{\rm NLO}_{t\bar{t} jj}
({\rm correlated})   &=& 0.0109^{+0.0043(39\%)}_{-0.0014(13\%)} \,, \nonumber \\ 
\nonumber \\
 \sigma^{\rm NLO}_{t\bar{t}b\bar{b}}/\sigma^{\rm NLO}_{t\bar{t} jj}
({\rm relative ~error})  &=& 0.0109^{+0.0026(24\%)}_{-0.0030(27\%)} \,.
\end{eqnarray}
After symmetrisation, the final theoretical errors amount to $32\%$
for the uncorrelated case and $26\%$ for the correlated one. Using the
relative error approach, we find $26\%$. Scale uncertainties for the
absolute $t\bar{t}b\bar{b}$ and $t\bar{t}jj$ cross sections are of the
order of $15\%-24\%$  ($14\%-21\%$ after symmetrisation) and therefore
comparable with the uncertainty  of the ratio. 

%
\subsection{LHC $@$ 13 TeV}
%
%
\begin{table}[h!]
\vspace{0.4cm}
\begin{center}
  \begin{tabular}{c|c|c|c}
&&&\\
$\sqrt{s}=$ 13 TeV & $\sigma^{\rm NLO}_{pp\to t\bar{t}b\bar{b}}$ [fb] &
$\sigma^{\rm NLO}_{pp\to t\bar{t} jj }$     [pb]   &
\textsc{Ratio} \\
&&&\\
\hline \hline
&&&\\
$\Delta R_{jj} > 0.8$, $R=0.8$&  886.8  $\pm$ 1.4  
& 86.7  $\pm$  0.1 & 0.0102 \\
&&&\\
\hline 
&&&\\
$\Delta R_{jj} > 0.5$, $R=0.5$ & 1078.3 $\pm$ 1.2  
& 85.5    $\pm$    0.2  & 0.0126\\
&&&\\
  \end{tabular}
\end{center}
\caption{\label{tab_13TeV}\it 
 NLO cross sections for $pp \to
  t\bar{t}b\bar{b} $ and  $pp \to t\bar{t} jj$ at the LHC with $\sqrt{s}$ = 13
  TeV, including errors from the Monte Carlo integration. The ratio
  $\sigma^{\rm NLO}_{pp\to t\bar{t}b\bar{b}}/\sigma^{\rm NLO}_{pp\to
    t\bar{t} jj }$ is  also shown. Results for two different values of
  jet resolution parameter $R$ and jets separation cut $\Delta R_{jj}$
  are reported.}
\end{table}
%
The case of $\sqrt{s}=13$ TeV shows a similar pattern. The NLO cross
sections for $t\bar{t}b\bar{b}$ and $t\bar{t} jj$ are reported in
Table \ref{tab_13TeV}, again for two different values of the jet
resolution parameter $R$ and jet separation cut $\Delta R_{jj}$. For
$\Delta R^{\rm min}_{jj}=R=0.5$ we find
\begin{equation}
\sigma^{\rm NLO}_{t\bar{t}b\bar{b}}
({\rm LHC_{13 TeV}},m_t=173.5 ~{\rm GeV,CT10})= 
1078.3^{+222.1(20\%)}_{-249.7(23\%)} ~{\rm [fb]}\,,
\end{equation}
\begin{equation}
\sigma^{\rm NLO}_{t\bar{t}jj}
({\rm LHC_{13 TeV}},m_t=173.5 ~{\rm GeV,CT10})= 85.5^{-18.3(21\%)}_{~-8.4(10\%)} 
~{\rm [pb]}\,.
\end{equation}
Scale uncertainties of the integrated cross sections are at the same
level as for $\sqrt{s}=$ 7 and 8 TeV and amount to $21\%-23\%$
($16\%-22\%$ after symmetrisation). The cross section ratio and its
estimated error amount to
\begin{eqnarray}
 \sigma^{\rm NLO}_{t\bar{t}b\bar{b}}/\sigma^{\rm NLO}_{t\bar{t} jj}
({\rm uncorrelated}) &=& 0.0126^{+0.0067(53\%)}_{-0.0029(23\%)} \,,  
  \nonumber \\ \nonumber \\
  \sigma^{\rm NLO}_{t\bar{t}b\bar{b}}/\sigma^{\rm NLO}_{t\bar{t} jj}
({\rm correlated}) &=& 0.0126^{+0.0067(53\%)}_{-0.0019(15\%)} \,, \nonumber \\
\nonumber \\
 \sigma^{\rm NLO}_{t\bar{t}b\bar{b}}/\sigma^{\rm NLO}_{t\bar{t} jj}
({\rm relative ~error})  &=& 0.0126^{+0.0037(29\%)}_{-0.0032(25\%)}\,.
\end{eqnarray}
For the uncorrelated case the theoretical error is $38\%$, whereas the
correlated approach gives $34\%$ and the relative-error approach
$27\%$. We observe here that the cross section ratio and its
uncertainty increases with the center-of-mass energy and the
difference between uncertainties evaluated in the correlated and
uncorrelated approaches becomes smaller. The theoretical error on the
ratio is in this case slightly larger than the corresponding one on
the absolute cross sections.

%
\begin{figure}
\begin{center}
\includegraphics[width=0.95\textwidth]{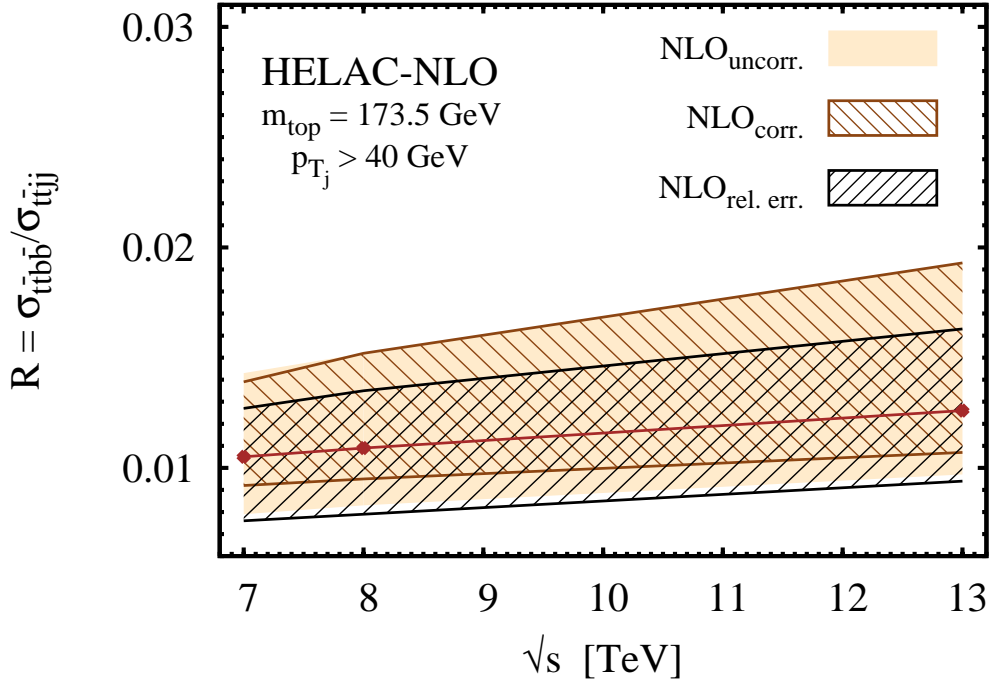}
\end{center}
\caption{\it \label{ratio-nlo} Theoretical  prediction for the
  $\sigma_{t\bar{t}b\bar{b}}/\sigma_{t\bar{t}jj}$  ratio at  the LHC
  as a function of the collider center-of-mass energy. Three
  uncertainty  bands correspond to  different methods of  estimating
  scale variations. }
  \end{figure}
%

We summarize our predictions in Figure \ref{ratio-nlo}, where the
cross section ratio is presented as a function of the collider
center-of-mass energy. The plot shows three different error bands
according to the three methods employed for the uncertainty
estimation. The error bands are  relatively independent on the method
adopted. The {\it uncorrelated} approach being the most conservative
one. We decided to adopt the latter for our comparison with the LHC
data at $\sqrt{s}=8$ TeV, that will be discussed in the next Section.

%
\section{Comparison with CMS Results at $\sqrt{s}=8$ TeV}
\label{comparison}
%
%
\begin{figure}
\begin{center}
\includegraphics[width=0.95\textwidth]{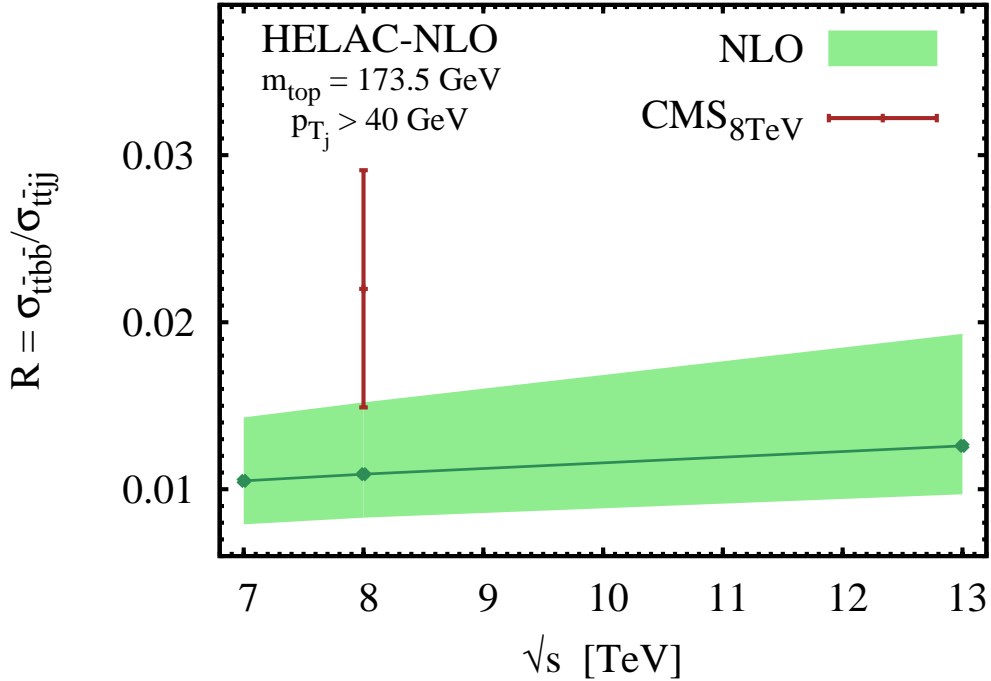}
\end{center}
\caption{\it \label{ratio-cms} Theoretical  prediction for the
  $\sigma_{t\bar{t}b\bar{b}}/\sigma_{t\bar{t}jj}$  ratio at  the LHC
  as a function of the collider center-of-mass energy, compared to
  the available measurement result from CMS  with $\sqrt{s}=8$ TeV. The 
  uncertainty  band depicts scale variation. }
  \end{figure}
%
We now compare our NLO predictions with the corresponding measurement
of the ratio $\sigma_{t\bar{t}b\bar{b}}/\sigma_{t\bar{t}jj}$ by the
CMS Collaboration \cite{CMS4,CMS6}, based on a data sample
corresponding to an integrated luminosity of $19.6 ~{\rm fb}^{-1}$
collected at $\sqrt{s} = $ 8 TeV in the di-lepton decay
mode. We quote below the experimental result, which has
been derived from the inclusive $t\bar{t}b\bar{b}$ and $t\bar{t}jj$ cross 
sections already unfolded to the full phase space for the top
quark.  In addition, the $t\bar{t}jj$ sample contains contributions from 
light-, charm- and bottom-jets.
Both rates take also into account the branching ratio of the
di-lepton decay mode of $10.8\%$ \cite{Beringer:1900zz} and
corrections  to the parton level jets. The result has been obtained for 
jets that are required to carry
$p_{T_{j}} > 20$ GeV ($p_{T_{j}} > 40$ GeV), to be located in the
rapidity range of $|y_{j}|<2.5$ and to be separated in the 
rapidity-azimuthal angle plane by  $\Delta
R_{jj}>0.5$:  
\begin{equation}
\sigma_{t\bar{t}b\bar{b}}/\sigma_{t\bar{t}jj}
({\rm LHC_{8 TeV}}, p_{T_{j}} > 20 ~{\rm GeV})=
0.021 \pm 0.003 ~{\rm (stat.)} \pm 0.005 ~{\rm (syst.)}\,,
\end{equation}
\begin{equation}
\sigma_{t\bar{t}b\bar{b}}/\sigma_{t\bar{t}jj}
({\rm LHC_{8 TeV}}, p_{T_{j}} > 40 ~{\rm GeV})=
0.022 \pm 0.005 ~{\rm (stat.)} \pm 0.005 ~{\rm (syst.)}\,.
\end{equation} 
A total systematic uncertainty of $22.6 \%$ has been estimated by CMS,
where the dominant contribution for $p_{T_{j}} > 40$ GeV comes from the
mistag rate ($12.6\%$) and the b-jet tagging efficiency ($11.2\%$)
\cite{CMS4}. Several experimental systematic uncertainties are
reduced by taking the cross section ratio, as expected. 

We can directly compare the measured ratio for $p_{T_{j}} > 40$ GeV with
the  corresponding \textsc{Helac-NLO} prediction at $\sqrt{s} = 8$
TeV:
\begin{equation}
\sigma^{\rm NLO}_{t\bar{t}b\bar{b}}/\sigma^{\rm NLO}_{t\bar{t} jj}
({\rm LHC_{8 TeV}},m_t=173.5 ~{\rm GeV,CT10})= 0.0109 +0.0043-0.0026 \,.
\end{equation} 
Let us remind that we have adopted here the most conservative {\it
  uncorrelated} approach for our   theoretical error estimate. As
Figure \ref{ratio-cms} also shows, our prediction calculated for the
central scale differs by a factor of 2 from the experimental
number. However, with the present level of accuracy the two results
agree within  $1.4\sigma$.
To facilitate the comparison, systematic
and statistical uncertainties reported by the CMS experiment have been
taken as uncorrelated and thus added in quadrature. The total
experimental error obtained in this way amounts at present to  
$\pm 0.0071$ ($32\%$). 

\begin{figure}
\begin{center}
\includegraphics[width=0.99\textwidth]{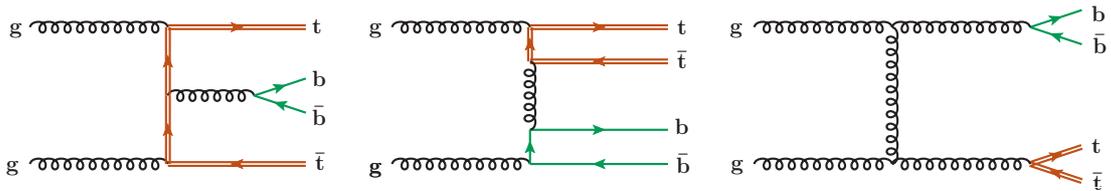}
\end{center}
\caption{\it \label{on-shell} Example of Feynman diagrams contributing
  to the  irreducible background $pp\to t\bar{t}b\bar{b}$  
at ${\cal O}(\alpha_s^4)$.}
\end{figure}

\begin{figure}
\begin{center}
\includegraphics[width=0.99\textwidth]{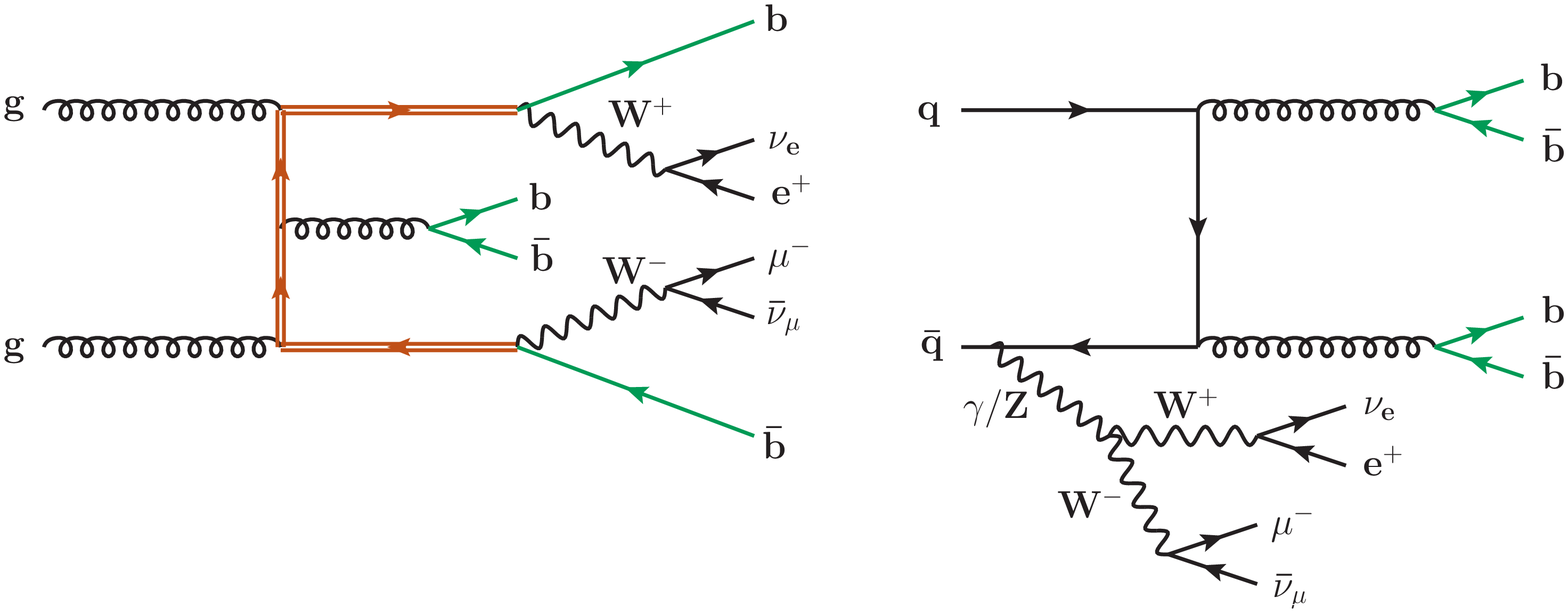}
\end{center}
\caption{\it \label{off-shell}  Example of Feynman diagrams
contributing to the  irreducible background  $pp\to e^{+}\nu_e
\mu^{-}\bar{\nu}_{\mu}b\bar{b} b\bar{b}$ at ${\cal{
O}}(\alpha_s^4\alpha^4)$ with  double-resonant (left panel) 
and non-resonant (right panel) top quark contributions. }
\end{figure}

  Let us conclude this section  by  discussing the impact
  of the top quark decays on the  $t\bar{t}b\bar{b}/t\bar{t}jj$
  ratio. To this end, we compare ratios evaluated with undecayed and
  decayed top quarks at the LO.  First  we have evaluated the ratio
  for $p_{T_{j}} > 40$ GeV, $|y_{j}|< 2.5$ and $\Delta R_{jj}>0.5$  at
  the LHC with $\sqrt{s}=8$ TeV using on-shell  top quarks for both
  processes {\it i.e.}  $\sigma^{\rm LO}_{t\bar{t}b\bar{b}}$  and
  $\sigma^{\rm LO}_{t\bar{t}jj}$ at ${\cal O}(\alpha_s^4)$ .
  In the next  step leptonic  decays
  of $W$ gauge bosons have been included.    More specifically $pp\to
  e^{+}\nu_e\mu^{-}\bar{\nu}_{\mu}b\bar{b}b\bar{b}$ and $pp\to
  e^{+}\nu_e\mu^{-}\bar{\nu}_{ \mu}b\bar{b}jj$ processes have been
  calculated with the help of \textsc{Helac-Phegas}, where  full
  off-shell and finite width top and W effects have been included  by
  taking  into account the double-resonant, single-resonant and
  non-resonant  contributions at order ${\cal O}(\alpha_s^4\alpha^4)$.
  Example of Feynman diagrams contributing to the LO $pp\to t\bar{t}
  b\bar{b}$  and $pp\to e^{+}\nu_e\mu^{-}\bar{\nu}_{
    \mu}b\bar{b}b\bar{b}$ processes are presented in Figure
  \ref{on-shell} and Figure \ref{off-shell}.   
 The following basic selection has been applied to
  (anti-)top decay products to ensure that the leptons are observed
  inside the detector and are well separated from each other:
\begin{equation}
p_{T_{\ell}} > 20 ~{\rm GeV}, ~~~|\eta_{\ell}| < 2.5,  ~~~\Delta
R_{\ell \ell}  > 0.4, ~~~\Delta R_{\ell j} > 0.4, ~~~p^{\rm miss}_{T}
> 30 ~{\rm GeV} \,,
\end{equation} 
where $p_{T}^{\rm miss}$ is  the transverse momentum of
the system of  two neutrinos.   The impact of the top quark decays on
the $\sigma_{t\bar{t}b\bar{b}}/\sigma_{t\bar{t}jj}$ cross section ratio has been
established to be less than $5\%$, well within the estimated dominant
theoretical uncertainties.

%
\section{Conclusions}
\label{conclusions}
%
%
In this paper we have presented the first consistent NLO theoretical
predictions for the cross section ratio $\sigma_{ttbb}/\sigma_{ttjj}$
in order to help high-quality comparisons with the data collected at
the LHC. We have considered the case of both present and future
collider energies, $\sqrt{s}=7$, $8$ and $13$ TeV, exploring different
methods to provide as much realistic estimates as possible of the
scale uncertainty for our predictions. We have found that our estimate
is relatively independent on the method applied. The method which
assumes $t\bar{t}b\bar{b}$ and $t\bar{t}jj$  uncorrelated should be
taken as the most conservative one. Moreover, we have shown that the scale
uncertainty of the ratio, at the level of $20\%-30\%$, is comparable
with the error on the absolute cross sections
$\sigma(t\bar{t}b\bar{b})$ and $\sigma(t\bar{t}jj)$. Given that this
uncertainty is the dominant theoretical error for the processes at
hand, we conclude that the ratio shows the same theoretical accuracy
as the individual cross sections \footnote{ We note, however, that
  ratios and double ratios of cross sections calculated at different
  center-of-mass energies could  lead to predictions with higher
  precision \cite{Mangano:2012mh}.}. 

Let us remind that top quark decays are not included in our
study. This corresponds to the unrealistic situation of a perfect top
quark reconstruction with all decay channels included. Besides, the
two light or $b$-jets are always assumed not to be misassociated with
the top quark decay products. It is clearly desirable to include top
quark decays and study how they affect the cross section ratio.
However, we expect a moderate impact, provided the same method of
reconstructing the $t\bar{t}$ system is used in both processes,
because the top quarks show a rather similar kinematics dependence in
$t\bar{t}b\bar{b}$ and $t\bar{t}jj$ backgrounds. This correlation
might be helpful to better distinguish whether the reconstructed
$b$-jets come from the $t\bar{t}$ pair, or {\it e.g.} from the QCD
$g\to b\bar{b}$ splitting. This assumption  is further
supported  by the study we have performed at the LO where very
moderate  effects  on the ratio coming from the top quark decays have 
been found.

The results presented in this paper have been obtained at the partonic level,
and parton shower effects should, in principle, be included. We expect
the parton
shower to play an important role in case of loose cuts on jet $p_T$,
{\it i.e.} for  $p_{T_{j}} \ll$ 40 GeV, where a question
mark is put on the reliability of a genuine fixed-order
calculation. First results for $t\bar{t}$ production in association
with up to two jets merged with parton shower have recently started to
appear \cite{Hoeche:2014qda}, but the assumed kinematical restriction
on the jet $p_T$ (40 GeV, 60 GeV or 80 GeV) 
seems still too high to shed light on such
effects. We note that the estimated uncertainty on the absolute cross
section for the production of  the $t\bar{t}jj$ system presented there
is comparable with our estimates. Similar conclusions apply as well to
the case of $t\bar{t}b\bar{b}$ production, recently matched to the
parton shower \cite{Cascioli:2013era}. 
Scale variations before and after matching have been
assessed to be rather similar, at the level of  $20\%-30\%$ which is
again in agreement with our estimates. Given all these reasons, we
believe that parton shower effects will have a minimal impact on our
results in the considered kinematical range.

Finally, we have presented a comparison between our NLO predictions
and the currently available CMS data for $\sqrt{s}=8$ TeV. The present
level of agreement is not striking but still within the
uncertainties. However, new measurements of the cross
  section ratio, based on the complete data samples collected by both
  the CMS and the ATLAS experiments, are
  underway. Those enlarged data samples,  including other top quark
  decay channels, will provide more accurate measurements, which we
  are looking forward to compare with our predictions.

%
\section*{Acknowledgments}
%
%

The authors would like to thank colleagues from the ATLAS and CMS
experiments for motivating them to perform this study. We acknowledge
Tae Jeong Kim for useful discussions, and Adam Kardos for
clarifications concerning parameter settings in \textsc{Pythia}.

The work of M. Worek was supported in part by the DFG under Grant
No. WO 1900/1-1 ({\it  "Signals and Backgrounds Beyond Leading
Order. Phenomenological studies for the LHC")}  and by the Research
Funding Program \textsc{Aristeta}, {\it "Higher Order Calculations and Tools for
High Energy Colliders"}, \textsc{HOCTools} (co-financed by the European Union
(European Social Fund ESF) and Greek national funds through the
Operational Program {\it "Education and Lifelong Learning"} of the National
Strategic Reference Framework (NSRF)).

\end{document}